\documentclass[a4paper,11pt]{article}
\pdfoutput=1
\usepackage{geometry}
\usepackage{soul}
\usepackage{a4wide,slashbox}
\usepackage{graphicx}
\usepackage{subfigure}
\usepackage{epsf}
\usepackage{amsmath}
\usepackage{amssymb}
\usepackage{caption}
\usepackage{subfig}
\usepackage{floatrow}

\usepackage{cite}
\usepackage{multirow,tabularx}
\usepackage{appendix}
\usepackage{tikz}
\usepackage{graphicx,amsmath,amsfonts,amssymb,amsthm,euscript,braket,xcolor}
\newcommand{\be}{\begin{equation}}
\newcommand{\ee}{\end{equation}}

\newcommand{\Rmnum}[1]{\expandafter\@slowromancap\romannumeral #1@}
\newcommand{\bea}{\begin{eqnarray}}
\newcommand{\eea}{\end{eqnarray}}

\numberwithin{equation}{section}

\begin{document}

\title{\bf On the time dependence of holographic complexity in a dynamical Einstein-dilaton model}
\author{\textbf{Subhash Mahapatra$^{a}$}\thanks{mahapatrasub@nitrkl.ac.in},
\textbf{Pratim Roy$^{b}$}\thanks{proy@niser.ac.in, pratroy@gmail.com},
 \\\\\
 \textit{{\small $^a$ Department of Physics and Astronomy, National Institute of Technology Rourkela, Rourkela - 769008, India}}\\
\textit{{\small $^b$ School of Physical Sciences, NISER, Bhubaneshwar, Khurda 752050, India}}}
\date{}

\maketitle
\abstract{
We study the holographic ``complexity=action'' (CA) and ``complexity=volume'' (CV) proposals in Einstein-dilaton gravity in all spacetime dimensions. We analytically construct an infinite family of black hole solutions  and use CA and CV proposals to investigate the time evolution of the complexity. Using the CA proposal, we find dimensional dependent violation of the Lloyd bound in early as well as in late times. Moreover, depending on the parameters of the theory, the bound violation relative to the conformal field theory result can be tailored in the early times as well. In contrast to the CA proposal, the CV proposal in our model yields results similar to those obtained in the literature.}

\section{Introduction}

Recent advances in the AdS/CFT correspondence \cite{Maldacena,Gubser:1998bc,Witten:1998qj} suggest non-trivial connections between different areas of physics, in particular between general relativity and quantum information. At the center of these developments is the seminal work of Ryu and Takayanagi \cite{Ryu:2006bv,Ryu:2006ef}, which provided a holographic dictionary for the calculation of entanglement entropy of the boundary theory. The Ryu-Takayanagi proposal, which states that the entanglement entropy of the boundary theory is equivalent to the area of a certain minimal surface in the bulk geometry subjected to some boundary conditions, is one of the most significant and fruitful ideas that has emerged from the AdS/CFT correspondence, providing not only a bridge connecting gravity with many body quantum systems but also providing a new tool in the study of quantum information theory. This idea has further been used to suggest that the dynamics of the bulk spacetime emerge from the quantum entanglement of the boundary theory \cite{VanRaamsdonk:2010pw,Faulkner:2013ica,Lashkari:2013koa}.

Aside from the entanglement entropy, another information theoretic quantity which is receiving a great attention of late is the quantum complexity \cite{Watrous,Aaronson}. Quantum complexity, which is a state dependent quantity, describes how many simple elementary gates are needed to obtain a particular state from some chosen reference state. Simply put, the complexity of a state $(B)$ with respect to a given initial state $(A)$ is defined as the least possible number of unitary transformations required to construct the state $B$ from $A$.  Although, quantum complexity in spin chain system has been studied in the literature \cite{Oliveira}, however the concept of complexity in quantum field theory is still uncharted territory. Only recently have complexity computations been developed for quantum field theories, see for example \cite{Chapman:2017rqy,Jefferson:2017sdb,Hashimoto:2017fga,Caputa:2017urj,Caputa:2017yrh,Caputa:2018kdj,Bhattacharyya:2018wym,
Bhattacharyya:2018bbv,Hackl:2018ptj,Khan:2018rzm,Yang:2018nda}. In the AdS/CFT correspondence, two potential holographic descriptions for the complexity have been suggested, namely, the complexity= action (CA) conjecture \cite{Brown:2015bva,Brown:2015lvg} and the complexity=volume (CV) conjecture \cite{Susskind:2014rva,Stanford:2014jda}.

A gravity system for which the holographic complexity has been studied extensively is the eternal two-sided $AdS$ black hole. This bulk geometry is dual to a thermofield double state in the dual boundary field theory,
\begin{equation}
\ket{\psi_{TFD}}  = \frac{1}{\sqrt{Z}} \sum_{j} e^{-E_j/(2T)} e^{-i E_j (t_L + t_R)} \ket{ E_{j} }_L \ket{ E_{j} }_R
\end{equation}
where $L$ and $R$ refer to the left and right regions of the two sided black hole which are entangled with each other.  The entanglement between the left and right copies of the boundary CFTs is due to the Einstein-Rosen bridge that connects the two regions. Since the complexity is conjectured to grow with time and this property is also shared with the Einstein-Rosen bridge, the authors of \cite{Susskind:2014rva, Stanford:2014jda} were led to conjecture that the complexity could be identified with the volume $\mathcal{V}$ of the maximal co-dimension one surface that ends at the boundary times $t_L$ and $t_R$ (see Fig. (\ref{fig:dCVdt})),
\begin{equation}
C_V = \frac{\text{max}(\mathcal{V})}{G_{d+1} \ell}
\label{CVproposal}
\end{equation}
where $G_{d+1}$ is the Newton's constant and $\ell$ is an arbitrary length scale. Eq. (\ref{CVproposal}) is the ``complexity=volume'' proposal for the holographic complexity. Using this proposal, it was shown that the time evolution of complexity for (d+1) dimensional AdS-Schwarzschild black hole satisfies the relation $\frac{dC_V}{dt}=8 \pi M/(d-1)$ in the late time limit. This holographic relation although can be analytically shown for planner AdS-Schwarzschild black holes, however numerically analysis is indeed to verify it for black holes with other horizon topologies, see for example \cite{Carmi:2017jqz}.

Another holographic proposal which has been put forward in the literature \cite{Brown:2015bva,Brown:2015lvg} is the ``complexity=action'' proposal. According to this proposal, holographic complexity is proportional to the bulk action evaluated in a certain spacetime region known as the Wheeler-De Witt patch,
\begin{equation}
C_A = \frac{S_{WDW}}{\pi}
\end{equation}
By construction this proposal is devoid of the ambiguity associated with arbitrary length scale $\ell$ which appears in the CV proposal \footnote{ The CA proposal however suffers from the ambiguities associated with the null joints and null boundary terms, which are presented in the total WDW action (see eq. (\ref{WDWaction})). We will discuss these ambiguities in more details in later sections.}. Interestingly, for AdS-Schwarzschild black holes, the late time behaviour of the rate of change of complexity using the CA proposal found to reach a constant value in all dimensions,
\begin{equation}
\frac{dC_A}{dt} \leq \frac{2M}{\pi}
\label{Lbound}
\end{equation}
where $M$ is the mass of the black hole. The above bound, which is often referred to as the ``Lloyd bound'', puts an upper bound on the rate of complexification of a system with mass $M$ \cite{lloyd2000ultimate}. The hope that the holographic complexity might provide useful information about the spacetime structure behind the horizon has led to intense investigation of the  CA and CV proposals in various gravity theories, probing their structure and properties
(see \cite{Lehner:2016vdi,Carmi:2016wjl,Chapman:2016hwi,Couch:2016exn,Brown:2017jil,Alishahiha:2017hwg,Cai:2016xho,Mazhari:2016yng,Momeni:2016ira,Pan:2016ecg,Cano:2018aqi,Barbon:2015ria,
Bolognesi:2018ion,Yang:2016awy,Reynolds:2016rvl,
Fu:2018kcp,Kim:2017lrw,HosseiniMansoori:2017tsm,Huang:2016fks,Agon:2018zso,Moosa:2017yvt,Bakhshaei:2017qud,Abad:2017cgl,
Alishahiha:2015rta,Ben-Ami:2016qex,Zangeneh:2017tub,Roy:2017kha,Roy:2017uar,Ageev:2018nye,Ling:2018xpc} for a necessarily incomplete selection of results). Interestingly,
it was found using the CA proposal that a large class of black holes have the same action growth rate at late times and therefore are conjectured to have the same complexity growth rate \cite{Alishahiha:2017hwg,Cai:2016xho,Pan:2016ecg,Cano:2018aqi}.

However, recent works in a variety of gravity systems have questioned the validity of the CA conjecture in eq. (\ref{Lbound}) and made it somewhat doubtful. In particular, by studying the time evolution of $\frac{dC_A}{dt}$, it was found that $\frac{dC_A}{dt}$ violates the Lloyd bound at early times, and then approaches the bound from above at late times \cite{Carmi:2017jqz} (see also \cite{Kim:2017qrq}). Motivated by the work of \cite{Carmi:2017jqz}, several other investigations found similar violation of the Lloyd bound in other gravity models as well \cite{HosseiniMansoori:2018gdu,An:2018xhv,An:2018dbz}. Recently, a few holographic systems, including Lifshitz and hyperscaling violating geometries, containing additional matter fields were also shown to exhibit the violation of Lloyd bound even at late times \cite{Couch:2017yil,Moosa:2017yiz,Swingle:2017zcd,Alishahiha:2018tep}.

In the light of above discussion its an important question to ask whether we can construct other gravity models or black hole solutions that can break the Lloyd bound. This may not only allow us to further constrain the validity of these holographic conjectures but also provide other examples in the growing list of literature where the Lloyd bound  can be explicitly violated, which in turn might help us to get a better understanding of the reason behind this violation in general holographic theories.

In this work, we further put the CA and CV proposals to test. We consider the Einstein-dilaton gravity system which has been extensively used in the AdS/CFT literature, for example to construct QCD like gauge theories (as the non-trivial profile of dilaton field, which generally corresponds to the running of the 't Hooft coupling constant in the dual field theory side, can break the conformal structure of the deep $IR$ region and provide a confinement behavior at low energies) from holography \cite{Karch:2006pv,Dudal:2017max,Dudal:2018ztm,Gursoy:2007er,Gursoy:2008za,Paula,He:2013qq,Yang:2015aia}. Such Einstein-dilaton models are indeed of great physical interest and it is relevant to see how incorporating a dilaton in the theory, thereby breaking the conformal symmetry, changes the physical behaviour of the complexity.

We first the solve the Einstein and dilaton equations of motion analytically in terms of an arbitrary scale function $A(z)$ (see eq. (\ref{ansatz})) in all spacetime dimensions and construct an infinite family of black hole solutions. We introduce two parameters $a$ and $n$ in our model via the scale function and work out the rate of change of complexity using both the CA and CV proposals. It turns out that, as opposed to the $AdS$-Schwarzschild case, the time evolution of $\frac{dC_A}{dt}$ depends explicitly on the spacetime dimension and is violated in both early as well as in late time limits in our model. The $\frac{dC_A}{dt}$ is still found to be proportional to the mass of the black hole, however the proportionality constant depends on the number of the spacetime dimension. In particular, the magnitude of the Lloyd bound violation is smaller for higher spacetime dimensions. Although the parameters $a$ and $n$ do not have much effects on the late time behavior of $\frac{dC_A}{dt}$, however they can effect its early time bound violation behaviour. Our model therefore provides another example where the Lloyd bound in holographic theories can be explicitly violated. Interestingly, in contrast to $\frac{dC_A}{dt}$, the time evolution of $\frac{dC_V}{dt}$ in our model does not lead to any violation of the results found in \cite{Stanford:2014jda,Carmi:2017jqz}.

The paper is organized as follows. In section \ref{EDsystem}, we introduce our Einstein-dilaton gravity model under investigation and obtain its analytic solution in all spacetime dimensions. In sections \ref{CAsection} and \ref{CVsection} we compute the rate of change of complexity using the two proposals. Finally, in section \ref{summary}, we summarize the main results and point out future directions.

\section{Einstein-Dilaton System}
\label{EDsystem}

In this section we will briefly discuss the gravity solution corresponding to the Einstein-dilaton system. We present only the relevant analytic expressions,
which will be important for our analysis later on, and we refer the reader to \cite{Dudal:2017max,Dudal:2018ztm} for more technical details relating to the derivation of the model.   The Einstein-dilaton system in $(d+1)$ dimensions is described by the action,
\begin{equation}
S = \frac{1}{16\pi G_{d+1}} \int d^{d+1}x \left[R - \frac{1}{2}(\partial \phi)^2 - V(\phi) \right] \,.
\label{action}
\end{equation}
where $G_{d+1}$ is the Newton constant in $d+1$ dimensions, $\phi$ is the dilaton field and $V(\phi)$ is the potential of the dilaton field $\phi$, whose explicit form will be described later on. Varying the above action, we get the following Einstein and dilaton equations
\begin{eqnarray}
& & R_{\mu \nu} -\frac{1}{2} R g_{\mu\nu} + \frac{1}{2} \biggl[ \frac{1}{2} g_{\mu\nu} (\partial \phi)^2 - \partial_{\mu}\phi \partial_{\nu}\phi + g_{\mu\nu}  V(\phi)  \biggr] = 0 , \nonumber \\
 & &  \partial_{\mu} \bigl[ \sqrt{-g} \partial^{\mu} \phi] - \sqrt{-g} \frac{\partial V}{\partial \phi} = 0 \,.
\end{eqnarray}
Interestingly, a complete analytic solution for the above gravity system can be found using the potential reconstruction technique in any dimension \cite{Dudal:2017max,Dudal:2018ztm} (see also \cite{Yang:2015aia}). In particular, using the below ans\"atze for the metric and dilaton field,
\begin{eqnarray}
 ds^2 &=& \frac{L^2 e^{2A(z)}}{z^2} \left( -g(z)dt^2 + \frac{dz^2}{g(z)} + d\vec{y}_{d-1}^2 \right) \ , \nonumber \\
 \phi &=& \phi(z) \,.
 \label{ansatz}
\end{eqnarray}
the Einstein and dilaton equations can be solved completely in terms of a single arbitrary scale factor $A(z)$. In general $d+1$ dimensions, we can solve for $g(z)$, $\phi(z)$ and $V(z)$ as,
\begin{eqnarray}
g(z) &=& 1- \frac{\int_{0}^{z} dx x^{d-1} e^{-(d-1) A(x)}}{\int_{0}^{z_h} dx x^{d-1} e^{-(d-1) A(x)}} \ , \nonumber  \\
\phi^{\prime}(z) &=& \sqrt{2(d-1) \biggl[A^{\prime}(z)^2 - A^{\prime \prime}(z) - \frac{2 A^{\prime}(z)}{z}\biggr]} \ , \nonumber \\
V(z) &=& -\frac{(d-1) z^2 g e^{-2A}}{L^2}  \left[A^{\prime \prime} + A^{\prime}\bigg((d-1)A^{\prime} -\frac{2(d-1)}{z}+\frac{3g^{\prime}}{2g}\biggr) \right. \nonumber \\ & &  \left.
 -  \frac{1}{z}\biggl(-\frac{d}{z}+\frac{3g^{\prime}}{2g}\biggr) + \frac{g^{\prime \prime}}{2(d-1)g} \right] \,.
\label{metsold}
\end{eqnarray}
where, $L$ denotes $AdS$ length, which shall henceforth be set to unity. To derive this solution, we have used the boundary condition that $g(0) = 1$ and $g(z_h)=0$, where $z_h$ is the horizon radius. In the following sections, it will be more convenient to work in the coordinate $r = \frac{1}{z}$, in which we may schematically write the metric and solution as,
\begin{eqnarray}
 ds^2 &=& -G(r)dt^2 + \frac{dr^2}{f(r)} + r^2 e^{2 A(r)} d\vec{y}_{d-1}^2 \, \nonumber  \\
g(r) &=& 1- \frac{\int_{r}^{\infty} dx \frac{ e^{-(d-1) A(x)}}{x^{d+1}}}{\int_{r_h}^{\infty} dx  \frac{e^{-(d-1) A(x)}}{x^{d+1}}} \ , \nonumber  \\
\phi^{\prime}(r) &=& \sqrt{2(d-1) \biggl[A^{\prime}(r)^2 - A^{\prime \prime}(r)\biggr]} \ , \nonumber \\
V(r) &=& -(d-1) r^2 g(r) e^{-2A(r)} \left[\frac{d}{r^2}+(d-1)A^{\prime}(r)^2 + A^{\prime \prime}(r) + A^{\prime}(r)\bigg( \frac{2d}{r}+\frac{3g^{\prime}(r)}{2g(r)}\biggr) \right. \nonumber \\ & &  \left.
 + \frac{(3d-1)}{2(d-1)}\frac{g\prime(r)}{r g(r)} + \frac{g^{\prime \prime}(r)}{2(d-1)g(r)} \right] \,.
\label{schematicmetric}
\end{eqnarray}
where $G(r)=e^{2A(r)}r^2 g(r)$ and $f(r) = e^{-2A(r)}r^2 g(r)$. We shall also record the expressions for the mass, temperature and the Bekenstein-Hawking entropy of the black hole,
\begin{eqnarray}
M &=& -\frac{V_{d-1}}{8 \pi G_{d+1}} \frac{r^{d+1}}{d} \left[ g'(r)+\frac{1}{2} r g''(r)   \right] \bigg\rvert_{r \rightarrow \infty}  \ , \nonumber \\
T &=& \frac{e^{-(d-1)A(r_h)}}{4\pi r_h^{d-1} \int_{0}^{1/r_h} dx x^{d-1} e^{-(d-1) A(x)}} \ , \nonumber \\
S &=& \frac{e^{(d-1)A(r_h)}r_h^{d-1}}{4 G_{d+1}} \,.
\label{MTS}
\end{eqnarray}
where in the above expressions we have used $r_h=\frac{1}{z_h}$.  The black hole mass expression in eq. (\ref{MTS}) is obtained using the Ashtekar-Magnon-Das (AMD) formulism, the details of which are relegated to Appendix A. \\

It is important to mention that eq. (\ref{schematicmetric}) is a solution of the Einstein-dilaton action for any scale factor $A(r)$. We therefore have an infinite family of analytic black hole solutions for the gravity system of eq. (\ref{action}). These different solutions however correspond to different dilaton potentials, as different forms of $A(r)$ will give different $V(r)$. Nonetheless, in the context of gauge/gravity duality, it is more reasonable to fix the form of $A(r)$ by taking inputs from the boundary theory. For example, in the area of holographic QCD \textit{i.e} $d=4$, the form of $A(r)$ is generally fixed by demanding physical properties such as confinement in the quark sector, linear Regge behaviour for the meson mass spectrum, confinement/decofinement transition etc. to be reproduced from holography. A particular interesting form of $A(r)=-a/r^n$ has been suggested in recent years in many works (see, for example \cite{Dudal:2017max,Dudal:2018ztm,Gursoy:2008za,Gursoy:2007er,Paula,He:2013qq,Yang:2015aia}) as it reproduces many of lattice QCD properties holographically. Moreover, one can also put a constraint on the value of $n$, such as $n>1$, by requiring confinement behaviour at low temperatures \cite{Gursoy:2007er}, and on $a$, such as $a>0$, by demanding the holographic confinement/deconfinement temperature transition to be around $270 / MeV$ in the pure glue sector \cite{Dudal:2017max}, as is observed in lattice QCD \cite{Fromm:2011qi}. In this work however, we will take a more general approach and consider different values of $n$ and $a$ (hence different dilaton potential) to see the effects of different running dilaton profiles on the time evolution of the holographic complexity in different dimensions. Note that we can change the strength of the dilaton fall-off by tuning the parameter $a$ and that this gives us a route to explicitly check the effect of the dilaton on the complexity growth.\\

It is easy to see that the scale factor $A(r)\rightarrow 0$ at the boundary $r=\infty$, asserting that spacetime asymptotes to AdS.  Moreover, for $a>0$, one can also note that near the asymptotic boundary
\begin{eqnarray}
& &V(r)|_{r\rightarrow \infty}=-\frac{(d-1)(d-2)}{L^2}+\frac{m^2 \phi^2}{2}+\ldots \,  \nonumber \\
& & V(r)|_{r\rightarrow \infty}=2 \Lambda + \frac{m^2 \phi^2}{2}+\ldots \,.
\label{Vcase1exp}
\end{eqnarray}
where as usual $\Lambda=-\frac{(d-1)(d-2)}{L^2}$ is the negative cosmological constant. $m^2$ is the mass of the dilaton field, which satisfies the Breitenlohner-Freedman bound for stability \textit{i.e}  $m^2\geq -d^2/4$ \cite{Breitenlohner:1982jf}. Moreover, the potential is bounded from above by its UV boundary value \textit{i.e} $V(\infty) \geq V(r)$. Therefore the gravity model under consideration also satisfies the Gubser criterion to have a well defined dual boundary theory \cite{Gubser:2000nd}. For this reason we will always consider $a>0$ case from on now, and case $a<0$ will be studied elsewhere.\\

For the calculations in the following sections, it is more convenient to write the metric in the Eddington-Finkelstein coordinates,
\begin{equation}
v=t+r^{*}(r) \ , ~~~~ u=t-r^{*}(r) \,.
\end{equation}
where $r^{*}(r)$ is defined as,
\begin{equation}
\label{tortoise}
dr^{*} = \frac{dr}{\sqrt{G(r)f(r)}} = \frac{dr}{r^2 g(r)} \,.
\end{equation}
Let us also note the expression of metric in these coordinate system as this will be useful later on,
\begin{eqnarray}
\label{Eddington-Finkelstein}
& & ds^2 = -G(r)dv^2 + 2e^{2A(r)}dvdr + r^2 e^{2A(r)} d\vec{y}_{d-1}^2 \ , \\ \nonumber
& & ds^2 = -G(r)du^2 - 2e^{2A(r)}dudr + r^2 e^{2A(r)} d\vec{y}_{d-1}^2 \ , \\ \nonumber
& & ds^2 = - G(r)dudv + r^2 e^{2A(r)} d\vec{y}_{d-1}^2 \,. \\ \nonumber
\end{eqnarray}

Before ending this section, it is important to mention that there also exists another admissible solution to the Einstein-dilaton equations of motion which corresponds to thermal $AdS$. This solution is obtained by taking the $r_h \rightarrow 0$ limit, which translates to $g(r)=1$. For $n>1$, we found the Hawking-Page type thermal-AdS/black hole phase transition between these two solutions. In particular, the black hole/thermal-AdS phases are found to be favoured at high/low temperatures respectively. Moreover, these thermal-AdS/black hole phases can be shown to be dual to confined/deconfined phases in the dual boundary theory (the detail investigation will appear elsewhere). On the other hand, for $n \leq 1$, the black hole solution is favored at all temperatures. In this work, in order to study the time dependence of holographic complexity, we will always consider the situation where the black hole phase is more stable. An interesting question, which we leave for future study, is to investigate how the time dependence of holographic complexity varies as we pass through the confinement/deconfinement critical point.

\section{Complexity using CA proposal}
\label{CAsection}

\begin{figure}[htp]
	\subfigure[]{
		\includegraphics[width=2.9in,height=2.4in]{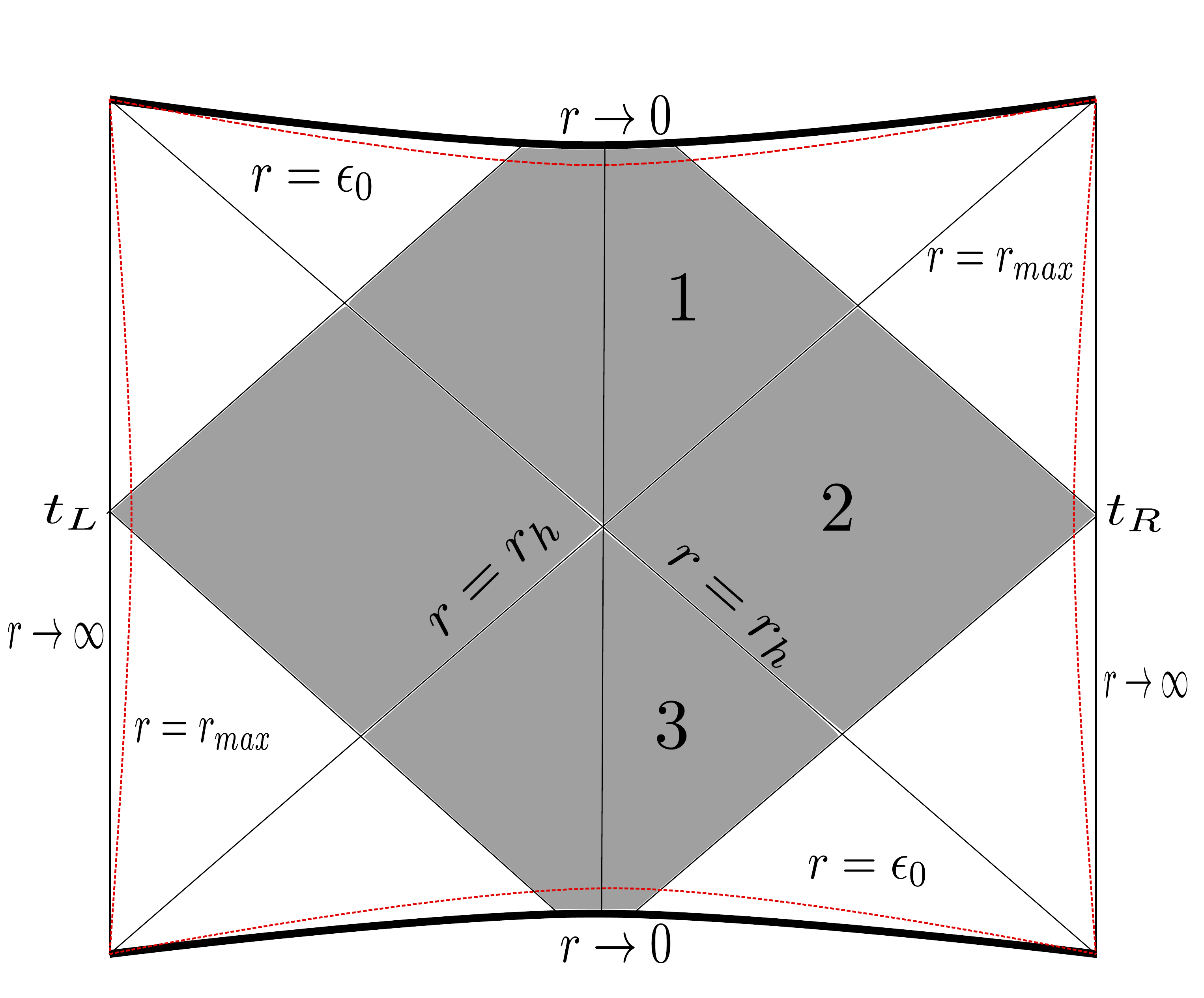}
	}
	\subfigure[]{
		\includegraphics[width=2.9in,height=2.4in]{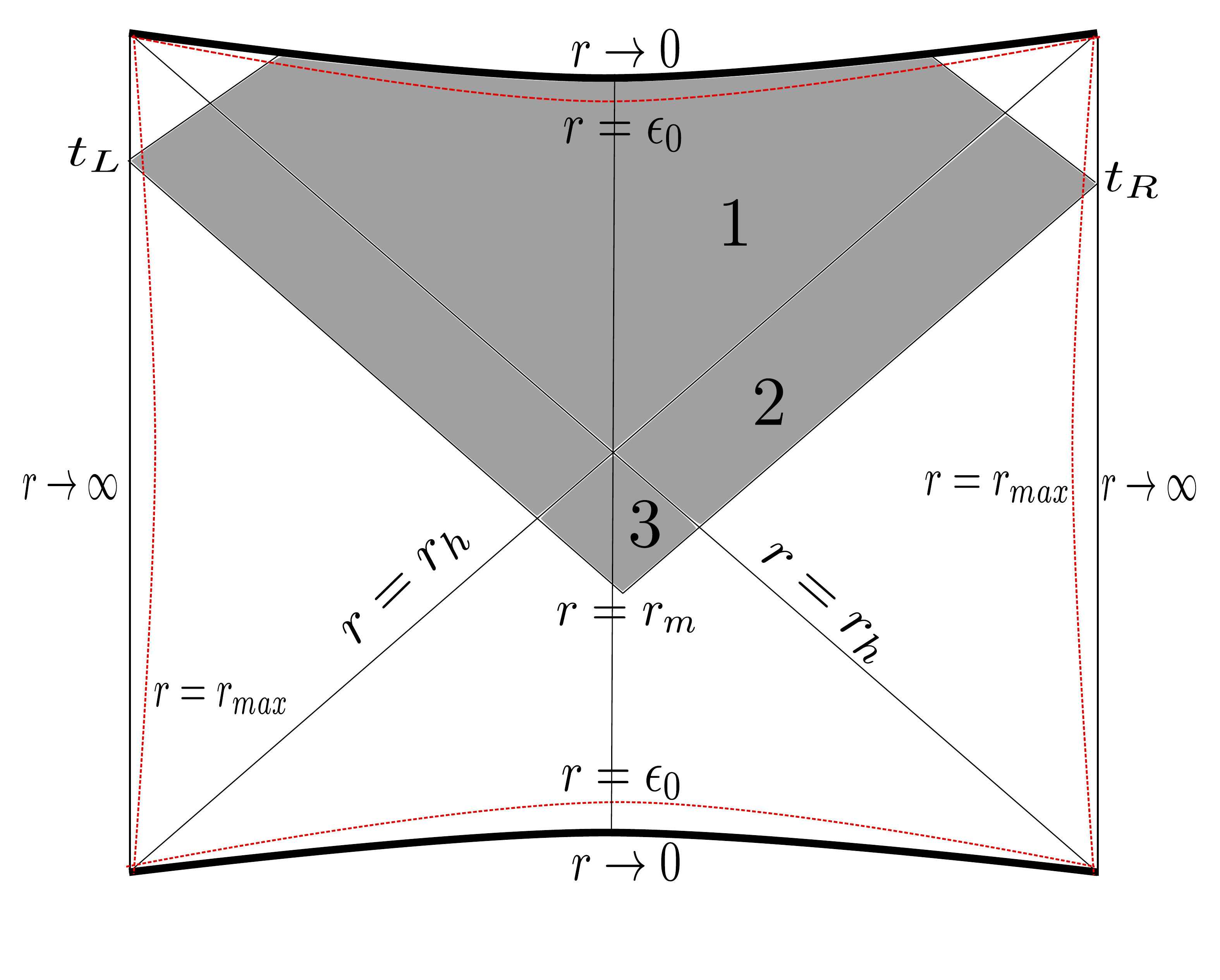}
	}
	\caption{Panel (a) shows the Penrose diagram at time $t<t_c$ and panel (b) shows the diagram at $t>t_c$.}
	\label{fig:wdw}
\end{figure}

In this section, we shall compute the complexity of the Einstein-dilaton system using the CA proposal \cite{Brown:2015bva} and the method developed in \cite{Lehner:2016vdi} for regions with null boundaries. As mentioned in the introductory section, the CA proposal consists of evaluating the action on a section of the spacetime known as the Wheeler-De Witt (WDW) patch. A WDW patch is defined in the following manner. Let us select two constant time slices $t_L$ and $t_R$ on the left and the right asymptotic boundaries respectively of an eternal black hole system. Let us also consider two null light sheets emanating from these two points. The patch is then defined as the region included between these light sheets and the points of intersection with the past and future singularities. This is encapsulated in Fig. (\ref{fig:wdw}).

As mentioned previously, we are interested in the time rate of change of complexity. By symmetry, this quantity would depend only on $t=t_L + t_R$ and not $t_L$ and $t_R$ individually. Furthermore, we henceforth consider a symmetric time evolution where $t=\frac{t_L}{2}=\frac{t_R}{2}$, as has also been done in \cite{Carmi:2017jqz}.\\

There are two possible situations that need to be taken into account to calculate the rate of change of complexity. Initially, the light sheets that delineate the WDW patch intersect the past singularity and then after a critical time $t_c$, the light sheets intersect with each other at a point $r=r_m$ without reaching the past singularity (Figures \ref{fig:wdw}(a) and (b) describe these situations). The critical time separating these two regimes can be calculated from the expression,
\begin{equation}
t_c = 2(r^{*}_{\infty} - r^{*}_{0}) \,.
\end{equation}

Now the full action that we require to evaluate the rate of change of complexity is given by \cite{Lehner:2016vdi},
\begin{eqnarray}
S_{WDW} &=& \frac{1}{16\pi G_{d+1}} \int d^{d+1}x \left[R - \frac{1}{2}(\partial \phi)^2 - V(\phi) \right]  \nonumber \\
 && +  \frac{1}{8\pi G_{d+1}} \int_{\mathcal{B}} d^{d} x \sqrt{|h|} K + \frac{1}{8\pi G_{d+1}} \int_\Sigma d^{d-1}x \sqrt{\sigma} \eta
 \nonumber \\
&& -\frac{1}{8\pi G_{d+1}} \int_{\mathcal{B}'}
d\lambda\, d^{d-1} \theta \sqrt{\gamma} \kappa
+\frac{1}{8\pi G_{d+1}} \int_{\Sigma'} d^{d-1} x \sqrt{\sigma} \ a \,.
\label{WDWaction}
\end{eqnarray}
The first line in the above expression is the familiar Einstein-dilaton action. The second term corresponds to the Gibbons-Hawking-York (GHY) surface contribution.  There will again be three surface contributions in our case - two coming from the spacelike surfaces at past and future singularity at $r=\varepsilon_{0}$, and one from the timelike surface at the asymptotic boundary $r=r_{max}$. As usual, these surface contributions are defined in terms of the trace of the extrinsic
curvature $K$.  The third is the Hayward joint term, that arises due to the intersection of two boundary segments, but will not play a major role here. The fourth term is the null boundary contribution defined in terms of a parameter $\kappa$ which measures the failure of the null generators to be affinely parametrized. Adopting the convention followed in \cite{Lehner:2016vdi}, we affinely parametrize the generators as a result of which we may set $\kappa=0$.  And the last term is the null joint contribution coming from the intersection of two surfaces, where at-least one of the surface is null. The explicit form of the null joint term depends on the precise nature of surfaces intersecting to form the joint and may be calculated according to the rules given in \cite{Lehner:2016vdi}.

 It is important to mention that the null boundary terms in eq. (\ref{WDWaction}), associated with null joints and null boundary surfaces, introduce certain ambiguities in the value of $S_{WDW}$. The influence of the these ambiguities on the holographic complexity was carefully studied in \cite{Lehner:2016vdi,Chapman:2016hwi}, where it was shown that these ambiguities do not affect the rate of change of holographic complexity. We have explicitly checked that this conclusion remains the same in our model as well.

To calculate the time dependence of holographic complexity we divide the Penrose diagram symmetrically into two parts: Right and left. These parts can further be divided into three regions 1, 2 and 3 (see Fig. \ref{fig:wdw}). We evaluate the relevant contributions to the action of the WDW patch coming from regions 1, 2 and 3 of the right part of the Penrose diagram and then simply multiply by a factor of two to account for contributions from the left part of the Penrose diagram.

\subsection{Time rate of change of holographic complexity for $t<t_c$}
It may be shown by arguments entirely similar to \cite{Carmi:2017jqz} that the holographic complexity is time independent when $t<t_c$ and so we address it only briefly here.\\
\\
The bulk contribution in the given geometry (\ref{schematicmetric}) evaluates to,
\begin{eqnarray}
S_{bulk} = \frac{V_{d-1}}{16\pi G_{d+1}} \int dr dt \left[\frac{2}{d-1}r^{d-1} e^{(d+1)A(r)}V(r)\right] \,.
\end{eqnarray}
As mentioned, we shall compute the above integral for the three regions given in the right side of Fig. \ref{fig:wdw}. In these regions, the expressions of the bulk contributions reduce to,
\begin{eqnarray}
S^1_{bulk} &=& \frac{V_{d-1}}{16\pi G_{d+1}} \int_{0}^{r_h}dr \left[\frac{2}{d-1}r^{d-1} e^{(d+1)A(r)}V(r) \right] \left(\frac{t}{2} + r^{*}_{\infty} - r^{*}(r) \right) \ ,  \nonumber \\
S^2_{bulk} &=& \frac{V_{d-1}}{8\pi G_{d+1}} \int_{r_h}^{\infty} dr \left[\frac{2}{d-1}r^{d-1} e^{(d+1)A(r)}V(r)\right] \left(r^{*}_{\infty} - r^{*}(r) \right) \ ,  \nonumber \\
S^3_{bulk} &=& \frac{V_{d-1}}{16\pi G_{d+1}} \int_{0}^{r_h}dr \left[\frac{2}{d-1}r^{d-1} e^{(d+1)A(r)}V(r) \right] \left(-\frac{t}{2} + r^{*}_{\infty} - r^{*}(r) \right) \,.
\label{bulktlesstcfull}
\end{eqnarray}
where $V_{d-1}$ is the volume of the spacelike directions of the boundary theory. The sum total of the above three contributions yields (including an extra factor of two to account for the left side of the Penrose diagram in Fig. \ref{fig:wdw}),
\begin{equation}
\label{bulktlesstc1}
S^{0}_{bulk} = \frac{V_{d-1}}{4\pi G_{d+1}} \int_{0}^{\infty} \left[\frac{2}{d-1}r^{d-1} e^{(d+1)A(r)}V(r)\right] \left(r^{*}_{\infty} - r^{*}(r) \right) \,.
\end{equation}
which is a time independent quantity and hence does not need to be taken into consideration for calculating the rate of change of complexity. \\
\\
We now evaluate the surface (GHY) contributions coming from the regulated surfaces at past and future singularities ($r=\epsilon_0$) and from the UV regulator surface at $r=r_{max}$. We first record the trace of the extrinsic curvature for the induced metric at a fixed $r$ surface. The expression is given by,
\begin{equation}
K=\nabla_{\mu} n^{\mu} = \pm \frac{1}{r^{d-1} e^{(d+1)A(r)}} \partial_{r} \left[ r^{d-1} e^{(d-1)A(r)} \sqrt{|G(r)|}\right] \,.
\label{extrinsic}
\end{equation}
where $+$ and $-$ signs are for $r=r_{max}$ and $r=\epsilon_0$ surfaces respectively. To obtain the second inequality in eq. (\ref{extrinsic}), we used the fact that the normals at $r=r_{max}$ and $r=\epsilon_0$ surfaces are given by,
\begin{eqnarray}
\textbf{n} = s_{\mu}dx^{\mu} &=& \frac{dr}{\sqrt{f(r_{max})}}, ~~~ r \rightarrow r_{max} \ , \\ \nonumber
          = t_{\mu}dx^{\mu} &=& \frac{-dr}{\sqrt{-f(r_{\epsilon_0})}}, ~~~ r \rightarrow \epsilon_0 \,.
           \label{normals}
\end{eqnarray}
Using eq. (\ref{extrinsic}), the GHY surface contributions from the past and future singularities and from the UV boundary are now given by,
\begin{eqnarray}
S^{past}_{surf} &=& -\frac{V_{d-1}}{8\pi G_{d+1}}  e^{-2A(r)} \sqrt{|G(r)|} \partial_{r} \left[r^{d-1} e^{(d-1)A(r)}\sqrt{|G(r)|} \right] \left(-\frac{t}{2} + r^{*}_{\infty} -r^{*}(r) \right)\bigg\rvert_{r=\epsilon_0} \ ,  \nonumber \\
S^{future}_{surf} &=& -\frac{V_{d-1}}{8\pi G_{d+1}}  e^{-2A(r)} \sqrt{|G(r)|} \partial_{r} \left[r^{d-1} e^{(d-1)A(r)}\sqrt{|G(r)|} \right] \left(\frac{t}{2} + r^{*}_{\infty} -r^{*}(r) \right)\bigg\rvert_{r=\epsilon_0} \ ,  \nonumber \\
S^{UV}_{surf} &=& \frac{V_{d-1}}{8\pi G_{d+1}}  e^{-2A(r)} \sqrt{|G(r)|} \partial_{r} \left[r^{d-1} e^{(d-1)A(r)}\sqrt{|G(r)|} \right] \left(r^{*}_{\infty} -r^{*}(r) \right) \bigg\rvert_{r=r_{max}} \,.
\end{eqnarray}
which can be further simplified to the following equations,
\begin{eqnarray}
S^{past}_{surf} &=& -\frac{V_{d-1} \ d }{16\pi G_{d+1}} r^{d+1} e^{(d-1)A(r)} \left[\frac{2 g(r)}{r}+\frac{g'(r)}{d}+2 g(r)A'(r) \right] \left(-\frac{t}{2} + r^{*}_{\infty} -r^{*}(r) \right)\bigg\rvert_{r=\epsilon_0} \ ,  \nonumber \\
S^{future}_{surf} &=& -\frac{V_{d-1} \ d}{16\pi G_{d+1}}  r^{d+1} e^{(d-1)A(r)} \left[\frac{2 g(r)}{r}+\frac{g'(r)}{d}+2 g(r)A'(r) \right] \left(\frac{t}{2} + r^{*}_{\infty} -r^{*}(r) \right)\bigg\rvert_{r=\epsilon_0} \ ,  \nonumber \\
S^{UV}_{surf} &=& \frac{V_{d-1} \ d}{16\pi G_{d+1}} r^{d+1}  e^{(d-1)A(r)} \left[\frac{2 g(r)}{r}+\frac{g'(r)}{d}+2 g(r)A'(r) \right] \left(r^{*}_{\infty} -r^{*}(r) \right) \bigg\rvert_{r=r_{max}} \,.
\label{surftlesstc11}
\end{eqnarray}
We observe that similarly to the bulk contribution, the UV surface contribution is also time independent and hence would not change at $t>t_c$ either. Moreover, one may also notice that the total surface contribution of the  remaining surfaces, which is now equal to
\begin{equation}
S^{0}_{surf} = -\frac{V_{d-1} d }{4\pi G_{d+1}} r^{d+1}  e^{(d-1)A(r)} \left[\frac{2 g(r)}{r}+\frac{g'(r)}{d}+2 g(r)A'(r) \right] \left(r^{*}_{\infty} -r^{*}(r) \right)\bigg\rvert_{r=\epsilon_0} \,.
\label{surftlesstcfull}
\end{equation}
is also independent of time and hence does not contribute to the time rate of change of complexity either. Therefore, the total contribution of the GYH surface terms to the rate of change of complexity is zero for $t \le t_c$ as well.\\
\\
Now we calculate the null joint contributions. There are a number of null joints, such as at the interaction of null boundaries of the WdW patch with the regulated surfaces at $r=r_{max}$ and $r=\epsilon_0$, that contribute in our case.  These contributions can be evaluated from the action \cite{Lehner:2016vdi},
\begin{eqnarray}
S_{jnt} &=& \frac{1}{8\pi G_{d+1}} \int_{\Sigma'} d^{d-1} x \sqrt{\sigma} \ a \,.
\end{eqnarray}
with
\begin{eqnarray}
a &=& -\text{sign}(k.t)\text{sign}(k.\hat{s}) \log{|k.t|}, \ \ \ \ \ \    \text{for spacelike-null joint} \\ \nonumber
 &=& -\text{sign}(k.s)\text{sign}(k.\hat{t}) \log{|k.s|},   \ \ \ \ \ \  \text{for timelike-null joint} \,.
\end{eqnarray}
where, $\hat{t}$ and $\hat{s}$ are two auxiliary unit vectors which are orthogonal to the timelike/spacelike-null junctions and are defined in the tangent space of the timelike/spacelike surfaces at $r=r_{max}$ and $r=\epsilon_0$. For more details on the notations and other technical issues at the null junctions, see \cite{Lehner:2016vdi}.\\

Now using $\hat{s}=\hat{s}^{\mu} \partial_{\mu}=\partial_{t}/\sqrt{-f(r)}$ in the case of spacelike-null joint (where the null WDW boundary meets the regulated surface $r=\epsilon_0$) and $\hat{t}=\hat{t}^{\mu} \partial_{\mu}=\partial_{\mu}/\sqrt{f(r)}$ in the case of timelike-null joint (where the null WDW boundary meets the $UV$ regulated surface $r=r_{max}$) we can easily evaluate the various null joint contributions
\begin{eqnarray}
\label{nulltlesstcfull}
S^{future}_{jnt} &=& -\frac{V_{d-1}}{16\pi G_{d+1}} r^{d-1} e^{(d-1)A(r)} \log{|G(r)|} \bigg\rvert_{r=\epsilon_0}  \ ,  \\ \nonumber
S^{UV}_{jnt} &=& \frac{V_{d-1}}{16\pi G_{d+1}} r^{d-1} e^{(d-1)A(r)} \log{|G(r)|} \bigg\rvert_{r=r_{max}}  \ ,   \\ \nonumber
S^{past}_{jnt} &=& -\frac{V_{d-1}}{16\pi G_{d+1}} r^{d-1} e^{(d-1)A(r)} \log{|G(r)|} \bigg\rvert_{r=\epsilon_0} \,.
\end{eqnarray}
We see from the above equation that the total as well as the individuals null contributions are time independent and hence do not effect the time rate of change of complexity. Therefore, combining eqs. (\ref{bulktlesstc1}), (\ref{surftlesstcfull}) and (\ref{nulltlesstcfull}) and noting that these are the only relevant terms contributing to the complexity at $t \le t_c$, we may thus conclude,
\begin{equation}
\frac{dC_A}{dt} = \frac{1}{\pi} \frac{S_{WDW}}{dt} = 0, \ \ \ \  \  \text{for} \ \ \  t \le t_c
\end{equation}
At this point it is interesting to recall that the same behaviour $\frac{dC_A}{dt} =0$ was obtained for $AdS$-Schwarschild black hole in \cite{Carmi:2017jqz}. Therefore we see that the introduction of dilaton field does not modify the complexification rate for $t<t_c$. As we will see shortly, the dilaton field does however change the complexity rate for $t>t_c$.

\subsection{Time rate of change of holographic complexity for $t>t_c$}

Now, we concentrate on the regime $t>t_c$ and systematically calculate the rate of change of WDW action,
\begin{equation}
\frac{dS}{dt} = \frac{dS_{bulk}}{dt} + \frac{dS_{surf}}{dt} + \frac{dS_{jnt}}{dt} \,.
\end{equation}
Fig. (\ref{fig:wdw}) reveals the change in the evaluation of $S_{WDW}$ when $t>t_c$ compared to $t<t_c$. The most obvious change is the inclusion of a new null-joint term due to the formation of a null-null joint at $r=r_m$. There are also some changes in the calculation of bulk and surface terms which are as follows. The bulk contributions are again divided into three regions 1, 2 and 3 as shown in the right panel (b) of Fig. (\ref{fig:wdw}). The contributions from regions $1$ and $2$ remain identical compared to eq. (\ref{bulktlesstcfull}), while the contribution from the region $3$ is now modified as,
\begin{eqnarray}
S^3_{bulk} &=& \frac{V_{d-1}}{16\pi G_{d+1}} \int_{r_m}^{r_h}dr \left[\frac{2}{d-1}r^{d-1} e^{(d+1)A(r)}V(r)\right] \left(-\frac{t}{2} + r^{*}_{\infty} - r^{*}(r) \right) \,.
\end{eqnarray}
Hence the total change in the bulk contribution compared to $t < t_c$ is given by,
\begin{equation}
\label{bulk}
\delta S_{bulk} = \frac{V_{d-1}}{8\pi G_{d+1}} \int_{0}^{r_m} dr \left[\frac{2}{d-1}r^{d-1} e^{(d+1)A(r)}V(r)\right] \left( \frac{\delta t}{2}+r^{*}(r)-r^{*}(0)\right) \,.
\end{equation}
where we have defined $\delta t = t - t_c$. At this point, let us recall that the specific form of $V(r)$ depends on the form of $g(r)$ in eq. (\ref{schematicmetric}). We find that it is possible to evaluate $g(r)$ and $V(r)$ in closed form for $A(r) = -a/r^n$ for any $n$. With $g(r)$ and $V(r)$ in hand, we may readily evaluate the above equation for $\delta S_{bulk}$. In principle, this may be done for all $n$, but in this paper we confine ourselves to $n=1, 2$, for which the evaluation is analytically tractable.\\

The next term that contributes to the complexity is the surface contribution due to the GHY term. For $t>t_c$, the contribution from the past singularity is absent and the contributions from the UV region and the future singularity retain the same form as given in (\ref{surftlesstc11}). Taking this into account, the total surface contribution to the change of action at $r=\epsilon_0$ is therefore given by,
\begin{eqnarray}
\label{surface}
\delta S_{surf} = \frac{2V_{d-1}}{8\pi G_{d+1}} e^{-2A(r)}\sqrt{-G(r)}\partial_{r} \left[r^{d-1} e^{(d-1)A(r)}\sqrt{-G(r)}\right] \left(-\frac{t}{2}+r^{*}_{\infty} - r^{*}(r) \right) \bigg\rvert_{r=\epsilon_0} \,.
\end{eqnarray}
\\

Similarly, the null-spacelike/timelike joint contributions that arise due to the interaction of null boundaries of the WdW patch with regulated surfaces at $r=r_{max}$ and $r=\epsilon_0$ (future) are again time independent and moreover have the same expressions as in $t\leq t_c$ case. Therefore they do not contribute to the rate of change of complexity for $t>t_c$ case either.\\

The only remaining contribution left to be computed is the null-null joint contribution that arises due to the intersection of the WdW patch null boundaries at $r=r_m$. This null-null joint term was absent and had no counterpart for $t<t_c$. According to the prescription given in \cite{Lehner:2016vdi}, the null-null joint term is given by,
\begin{equation}
S_{jnt}^{r=r_m} = \delta S_{jnt}^{r=r_m} = \frac{1}{8\pi G_{d+1}} \int dx_i \log \left( -\frac{1}{2}k.\bar{k} \right) \,.
\end{equation}
where $k$ and $\bar{k}$ are the null normals to the $v$ and $u$ surfaces and are given by,
\begin{eqnarray}
k &=& c \partial_{\mu} (t+r^{*}) \ , \\ \nonumber
\bar{k} &=& \bar{c} \partial_{\mu} (t-r^{*}) \,.
\end{eqnarray}
Using the above expressions and eq. (\ref{tortoise}) the joint contribution evaluates to,
\begin{eqnarray}
\label{joint}
\delta S_{jnt}^{r=r_m} &=& -\frac{V_{d-1} }{8\pi G_{d+1}} r_m^{d-1} e^{(d-1)A(r_m)}\log \left(\frac{G(r_m)}{c \bar{c}} \right) \,.
\end{eqnarray}
 We would like to again mention that this joint contribution is sensitive to the ambiguities associated with null joints i.e., through its dependence on the normalization constant $c$. However, as shown in \cite{Lehner:2016vdi,Chapman:2016hwi}, these ambiguities do not affect the time rate of change of holographic
complexity. In any case, we have explicitly checked that different values of $c,\bar{c}$ do not qualitatively change our results for the holographic complexity.

It should be noted that both $\delta S_{bulk}$ and $\delta S_{jnt}^{r=r_m}$ are implicitly time dependent since $r_m \equiv r_m(t)$. To obtain the functional dependence of $r_m$ on $t$, let us first note that
\begin{equation}
\label{deltatrm}
\frac{\delta t}{2} + r^{*}(r_m) - r^{*}(0) = 0 \,.
\end{equation}
which can be used to obtained the time dependence of $r_m$ as,
\begin{equation}
\label{drmdt}
\frac{d r_m}{dt} = -\frac{r_m^2 g(r_m)}{2} \,.
\end{equation}
In principle, eqs. (\ref{bulk}), (\ref{surface}), (\ref{joint}) and (\ref{drmdt}) constitute all the ingredients we needed to study the rate of change of holographic complexity for $t>t_c$. These equations, which are written in terms of scale factor $A(r)$, are very general and will be of same form for any Einstein-dilaton gravity with metric as in eq. (\ref{metsold}). However, the extended expressions of $\delta S_{bulk}$ etc depend non-trivially on $n$ (recall that $A(r)=-a/r^n$) and it may not be possible to write them in general form containing $n$. For this reason, we will take $n=1,2$ in the remainder of this section and explicitly evaluate the rate of change of complexity for these values of $n$.

\subsubsection{Case 1: $d=4$, $n=1$}

Let us first evaluate the three relevant terms for $n=1$ in $AdS_{5}$. For this purpose, let us first note the expression of $g(r)$,
\begin{equation}
\label{gd4n1}
g(r)  =  1-\frac{e^{\frac{3 a}{r}} \left(\frac{3 a \left(3 a^2-3 a r+2
   r^2\right)}{r^3}-2\right)+2}{e^{\frac{3 a}{r_h}} \left(\frac{3 a
   \left(3 a^2-3 a r_h+2 r_{h}^2\right)}{r_{h}^3}-2\right)+2} \,.
\end{equation}
Using eqs. (\ref{bulk}), (\ref{surface}) and (\ref{joint}) along with (\ref{deltatrm}), we get the following expressions,
\begin{eqnarray}
\label{n1d4full}
& \frac{d S_{bulk}}{dt} = \frac{-V_3}{8\pi G_5}\frac{r_m^2\left[ r_h^3 \left(3a^2 -4a r_m + 2 r_m^2 \right) + e^{3a \left(\frac{1}{r_h}-\frac{1}{r_m}\right)}r_m (a+r_m)(9a^3-9a^2 r_h+6a r_h^2-2 r_h^3) \right]}{9 a^3 e^{\frac{3 a}{r_h}}-9 a^2 r_h e^{\frac{3 a}{r_h}}-2 r_h^3 \left(e^{\frac{3 a}{r_h}}-1\right)+6 a r_h^2 e^{\frac{3 a}{r_h}}} \ ,  \nonumber \\
& \frac{d S_{surf}}{dt} = \frac{V_3}{16\pi G_5}\frac{45 a^4 r_h^3}{9 a^3 e^{\frac{3 a}{r_h}}-9 a^2 r_h e^{\frac{3 a}{r_h}}-2 r_h^3 \left(e^{\frac{3 a}{r_h}}-1\right)+6 a r_h^2 e^{\frac{3 a}{r_h}}} \ , \\
& \frac{d S_{jnt}}{dt} = \frac{V_3}{16\pi G_5}\frac{r_m^2 e^{-\frac{3 a}{r_m}} \left(1-\frac{e^{\frac{3 a}{r_m}} \left(\frac{3 a \left(3 a^2-3 a r_m+2 r_m^2\right)}{r_m^3}-2\right)+2}{e^{\frac{3 a}{r_h}} \left(\frac{3
   a \left(3 a^2-3 a r_h+2 r_h^2\right)}{r_h^3}-2\right)+2}\right) \left(3 r_m (a+r_m) G(r_m) \log \left(\frac{G(r_m)}{\alpha ^2}\right)+r_m^3
   G'(r_m)\right)}{G(r_m)} \,. \nonumber
\end{eqnarray}

Before going on to study the full time dependence of the holographic complexity, let us first investigate its late time behavior.  We note that for large $\delta t$ we have $r_m \rightarrow r_h$. This can be noted by performing a late time expansion of eq. (\ref{deltatrm}) as follows. Firstly we require evaluation of $r^{*}(r)$. Unfortunately, this can be done only approximately for our model. To first order in $a$, we have the expression,
\begin{equation}
r^{*}(r) = \frac{1}{2r_h} \tan^{-1} (r/r_h) + \log \left(\frac{|r-r_h|}{r+r_h} \right) + \frac{a}{4r_h^3}\left(\frac{2r r_h}{r^2+r_h^2} +4 \tan^{-1}(r/r_h) + \log \left(\frac{r+r_h}{|r-r_h|}\right) \right) \,.
\label{rsexp}
\end{equation}
Inserting into eq. (\ref{deltatrm}) and solving for $r_m$ we get,
\begin{equation}
 r_m = r_h \left(1-2e^{-\frac{\pi r_h^2+2a(1+\pi)+4r_h^3 \delta t}{8r_h^3+2a}} \right) + \mathcal{O}(r_h-r_m)^2 \,.
\end{equation}
As can be seen from the above equation $r_m \rightarrow r_h$ in the late time limit, which is also physically expected and has been observed in many cases in the literature before. We also like to explicitly mention here that we have checked $r_m$ approaches $r_h$ in the late time limit even when higher order expansions in $a$  are considered in eq. (\ref{rsexp}). Now, taking the limit $r_m \rightarrow r_h$ in eq. (\ref{n1d4full}), we get considerably simpler forms for the bulk and joint contributions,
\begin{eqnarray}
& & \frac{d S_{bulk}}{dt} = -\frac{V_3}{16 \pi G_5}\frac{18 a^4 r_h^3}{e^{\frac{3 a}{r_h}} \left(9 a^3-9 a^2 r_h+6 a r_h^2-2 r_h^3\right)+2 r_h^3} \ , \\ \nonumber
& & \frac{d S_{jnt}}{dt} = \frac{V_3}{16 \pi G_5} \frac{27 a^4 r_h^3}{e^{\frac{3 a}{r_h}} \left(9 a^3-9 a^2 r_h+6 a r_h^2-2 r_h^3\right)+2 r_h^3} \,.
\end{eqnarray}
Hence, in the late time limit, the total rate of change of complexity is given by,
\begin{equation}
\label{CAmaster}
\frac{d S}{dt} = \frac{V_3}{16\pi G_5} \frac{54 a^4 r_h^3}{e^{\frac{3 a}{r_h}} \left(9 a^3-9 a^2 r_h+6 a r_h^2-2 r_h^3\right)+2 r_h^3} \,.
\end{equation}
Let us also note the expression of black hole mass\footnote{See Appendix A for the calculation of black hole mass in Einstein-dilaton gravity in various dimensions.},
\begin{equation}
\label{massn1}
M =\frac{V_3}{64 \pi  G_5} \frac{81 a^4 r_h^3}{ \left(e^{\frac{3 a}{r_h}} \left(9 a^3-9 a^2 r_h+6 a r_h^2-2 r_h^3\right)+2 r_h^3\right)} \,.
\end{equation}
Using the preceding equations, we get the following relation between the mass of the black hole and the rate of change of complexity at late times,
\begin{equation}
\label{CAd5relation}
\lim_{t \rightarrow \infty} \frac{dC_A}{dt} =  \frac{8M}{3\pi} \,.
\end{equation}
which clearly violates the Lloyd bound $\frac{dC_A}{dt}\leq 2M/\pi$.

Having discussed the late time behavior of holographic complexity, we now proceed to discuss its full time dependence. Unfortunately, this can be done only numerically in our model. To this end, we observe that time dependence enters into above equations through the quantity $r_m \equiv r_m(t)$, whose variation with respect to $t$ is given by eq. (\ref{drmdt}). To find $r_m(t)$, we simply integrate eq. (\ref{drmdt}) with the initial condition $r_m(0) = \epsilon_0$. Having done so, we may easily evaluate $\frac{dC_A}{dt}$ by substituting $r_m(t)$ into eq. (\ref{n1d4full}).

\begin{figure}[ht]
	\subfigure[]{
		\includegraphics[scale=0.5]{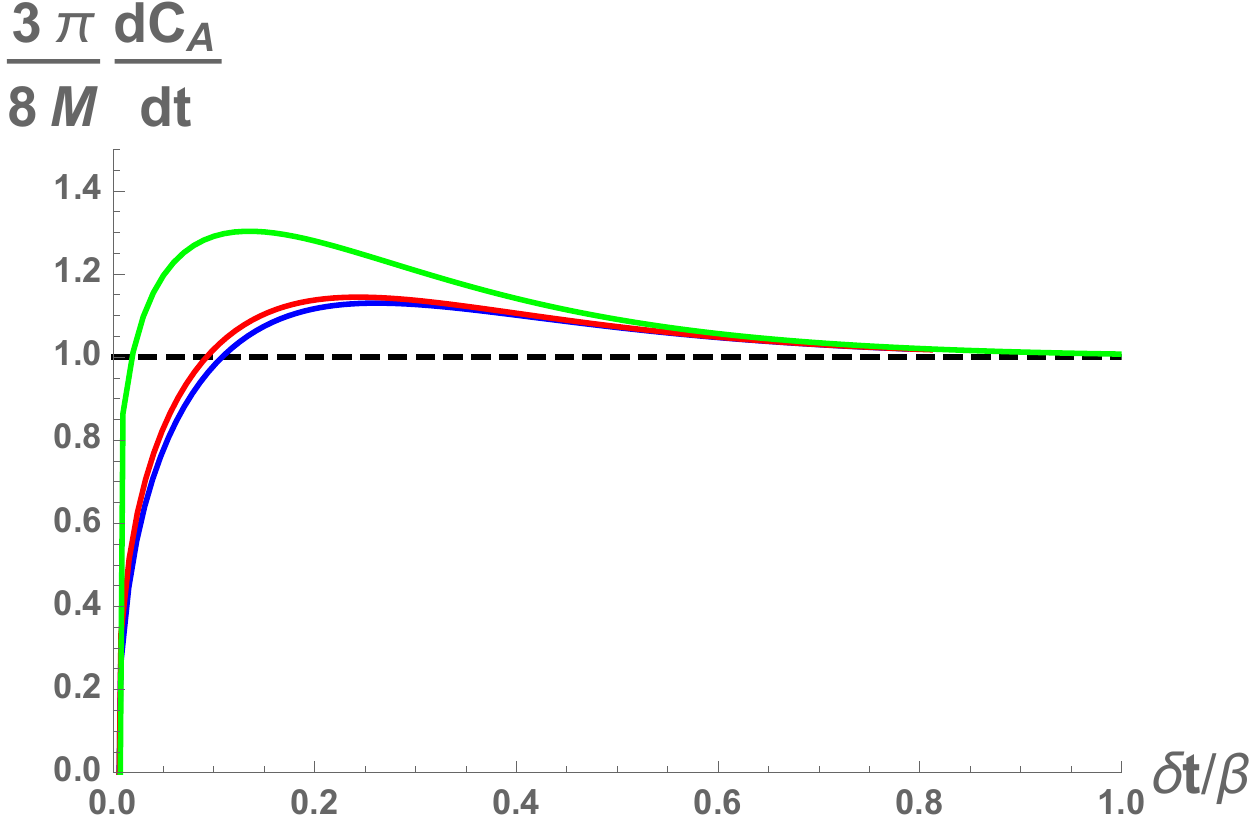}
	}
	\subfigure[]{
		\includegraphics[scale=0.5]{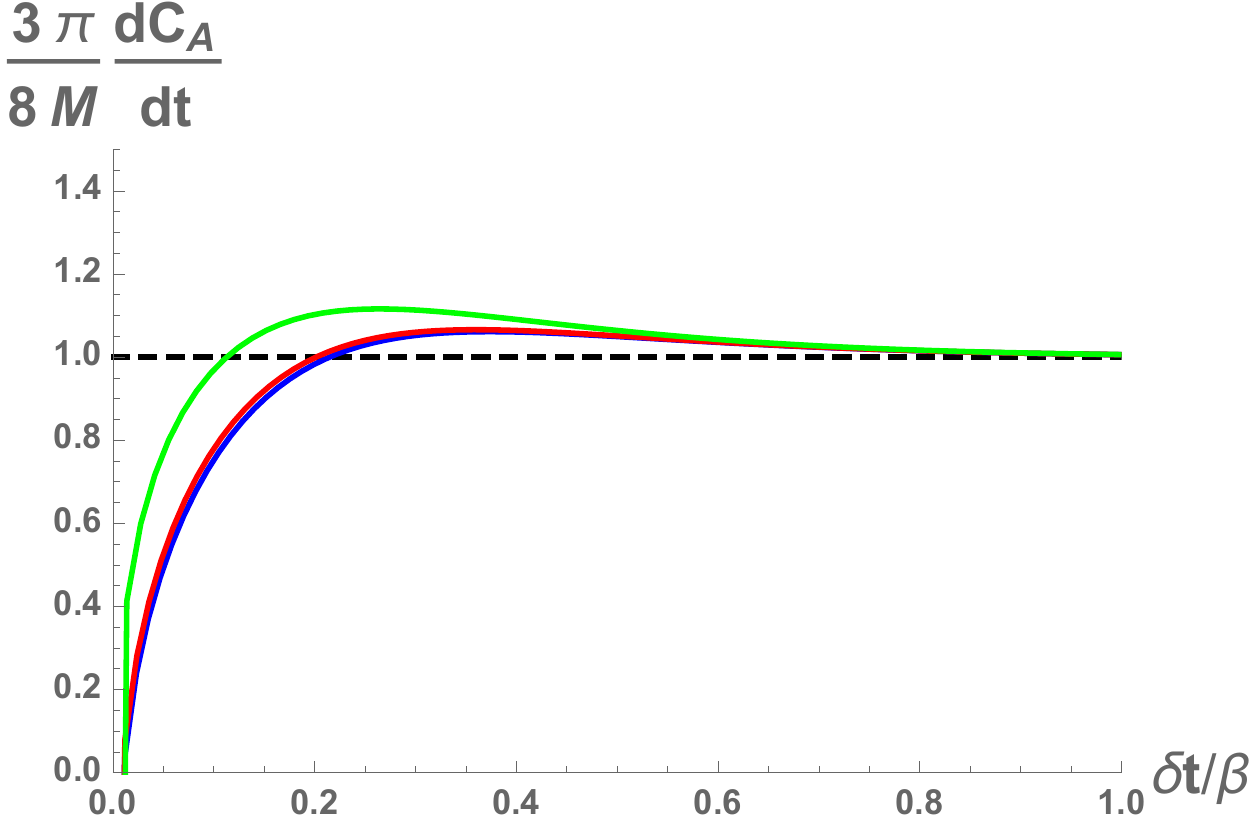}
	}
	\subfigure[]{
		\includegraphics[scale=0.5]{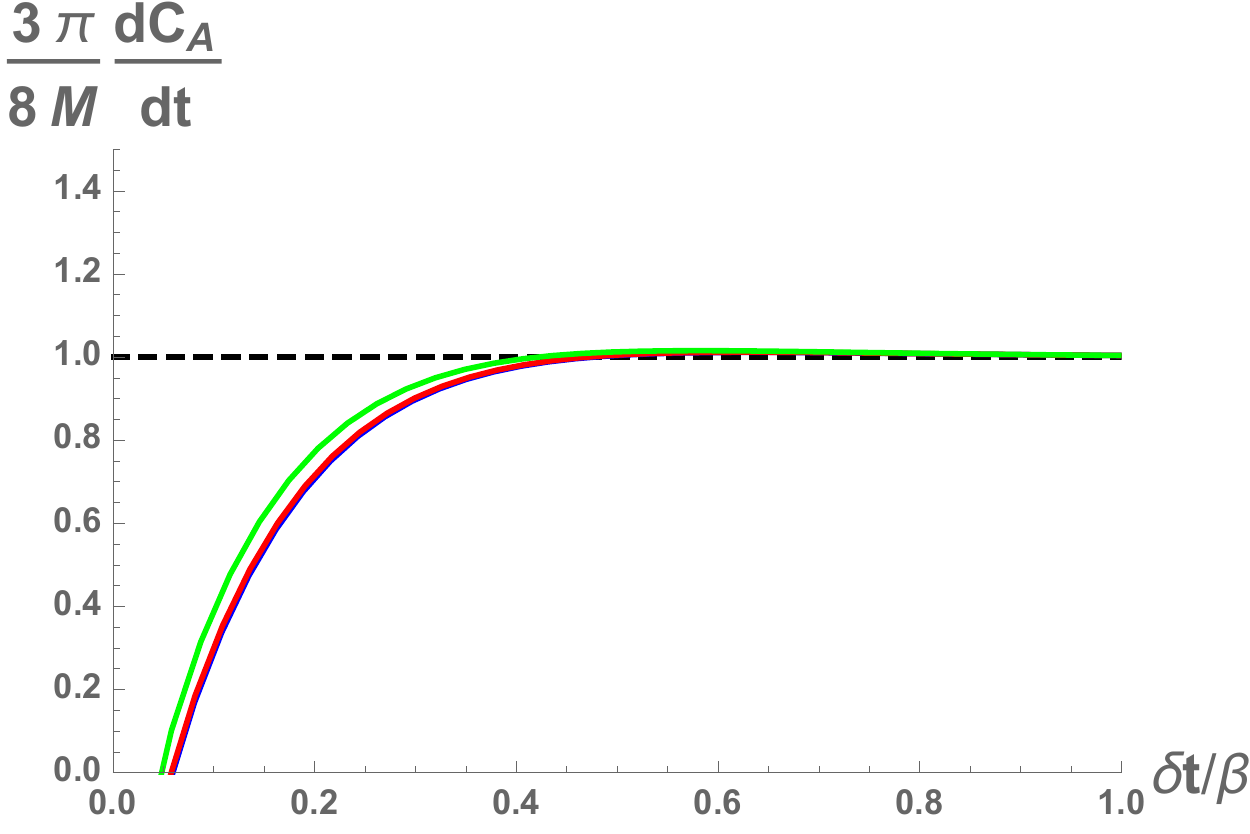}
	}
	\caption{$\frac{dC_A}{dt}$ is plotted against the dimensionless quantity $\delta t/\beta$  for $d=4$ and $n=1$. Here $\beta$ is inverse temperature. Panel (a) shows the variation of $\frac{dC_A}{dt}$ for $r_h = 1.0$, (b) for $r_h=1.5$ and (c) for $r_h = 3.5$.  In each panel blue, red and green curves correspond to $a=0.05$, $a=0.1$ and  $a=0.5$ respectively. In all cases, the dotted line represents the ratio $\frac{3 \pi}{8 M}\frac{dC_A}{dt}=1$. The values of the constants $c, \bar{c}=1$ have been set for simplicity.}
	\label{fig:dCAdtd4n1}
\end{figure}

The results for the time dependence of holographic complexity are shown in Fig. \ref{fig:dCAdtd4n1}, where three different values of $r_h$ are considered. Moreover, in order to make our analysis more complete, we have considered three different values of the parameter $a$ as well. It may be readily observed that at late times the rate of change of complexity asymptotes to $\frac{8 M}{3 \pi}$ in each case, as was also observed using the analytical calculations above. From the plot, we can also observe that the rate of change of complexity approaches its asymptotic value from above. The same result was also obtained in \cite{Carmi:2017jqz}, however with a asymptotic value that instead satisfied the Lloyd bound ($2M/\pi$). At this point, it is instructive to point out the main difference we found compared to the results of \cite{Carmi:2017jqz}. In particular, we found that  the Lloyd bound gets violated at all times (early as well as late) in our model whereas in \cite{Carmi:2017jqz} Lloyd bound get violated only at early times.

It is important to note here that we find numerical evidence of a negative peak at $\delta t = 0$. Such a negative peak was also analytically seen for $AdS$-Schwarzschild black hole in \cite{Carmi:2017jqz}, where $dC_A/dt \propto \log{\delta t}$ behavior at $\delta t = 0$ was found. Unfortunately, we are unable to analytically establish such a relation in our model. In this regard, in the current work we will present our numerical results for the time evolution of $dC_A/dt$ for the values of $\delta t$ slightly greater than zero.

Moreover, a new feature that appears in our model is that the magnitude of this early time bound violation from its late time value get increases with parameter $a$. This behavior is true irrespective to the value of $r_h$, as can be seen from Fig. \ref{fig:dCAdtd4n1}. It may be observed that for $a=0$, the results should be the same as $AdS$-Schwarzschild. However, this is manifestly not the case here. To determine the reason, let us consider the late time bulk contribution in eq. (\ref{n1d4full}). Applying the $a \rightarrow 0$ limit in this case yields,
\begin{equation}
 \lim_{a \rightarrow 0} \frac{d S_{bulk}}{dt} = -\frac{V_3}{16 \pi G_5}\frac{8 r_m^4}{3}
\end{equation}
whereas for $AdS$-Schwarzschild (see \cite{Lehner:2016vdi}), the bulk contribution instead gives
\begin{equation}
 \frac{d S_{bulk}}{dt} = - \frac{V_3}{16 \pi G_5} 2 r_m^4
\end{equation}
However, taking the limit $a \rightarrow 0$ on the integrand in eq. (\ref{bulk}) yields the expression,
\begin{equation}
 \lim_{a \rightarrow 0} \left[ \frac{V_3}{16 \pi G_5} \frac{2}{d-1}r^{d-1} e^{(d+1)A(r)}V(r)\right]\bigg\rvert_{d=4} = - \frac{V_3}{16 \pi G_5} 8 r^3
\end{equation}
which is exactly the expression one should obtain for $AdS$-Schwarzschild. Thus, we find that taking the limit $a \rightarrow 0$ and performing the integral do not commute in our model. This non-commuting nature between integral and $a \rightarrow 0$ limit is main reason for obtaining different results for $\frac{dC_A}{dt}$ in our model, and may be shown to be true for the boundary contribution as well. Curiously, the joint term does not exhibit the same feature, possibly because it does not require radial integration, as well taking the $r \rightarrow 0$ limit. Let us proceed to explain the $a \rightarrow 0$ limit in more detail. The expression for the dilaton field is,
\begin{eqnarray*}
	\phi(z) & = & \frac{\sqrt{2} z^{1-\frac{n}{2}} \sqrt{a (d-1) n z^{n-2}   \left(a n z^n+n+1\right)}}{n^{3/2}
		\sqrt{a \left(a n z^n+n+1\right)}} \times \nonumber \\ & & \left[\sqrt{a n} z^{n/2} \sqrt{a n z^n+n+1}+(n+1) \log \left(a n z^{n/2}+\sqrt{a n} \sqrt{a n z^n+n+1}\right)\right]
\end{eqnarray*}
It may easily be checked that putting $a < 0$ in the above expression results in an imaginary dilaton, which is physically not desirable, whereas the dilaton remains real for $a > 0$. Hence, $a<0$ is quite a different physical regime compared to $a>0$ in our case. Due to the absence of analytic continuation to $a<0$,
\begin{equation*}
\lim_{a \rightarrow 0^{+}} \frac{dS_{bulk}}{dt} \neq \lim_{a \rightarrow 0^{-}} \frac{dS_{bulk}}{dt}
\end{equation*}
This is the reason for the interesting $a \rightarrow 0$ limit in our case. We stress however, that the metric is perfectly valid as a solution of Einstein equations for $a>0$ and indeed one can take the value of $a$ to be arbitrarily small in our calculations.

We further find, a feature which is in common with \cite{Carmi:2017jqz}, that the deviation of $\frac{dC_A}{dt}$ in early times from its late time bound gets suppressed for larger values of $r_h$. Therefore the early time violation in $\frac{dC_A}{dt}$ gets washed out for higher temperatures/larger radius black holes in our case as well.

\begin{figure}
	\begin{tabular}{cc}
		\includegraphics[scale=0.4]{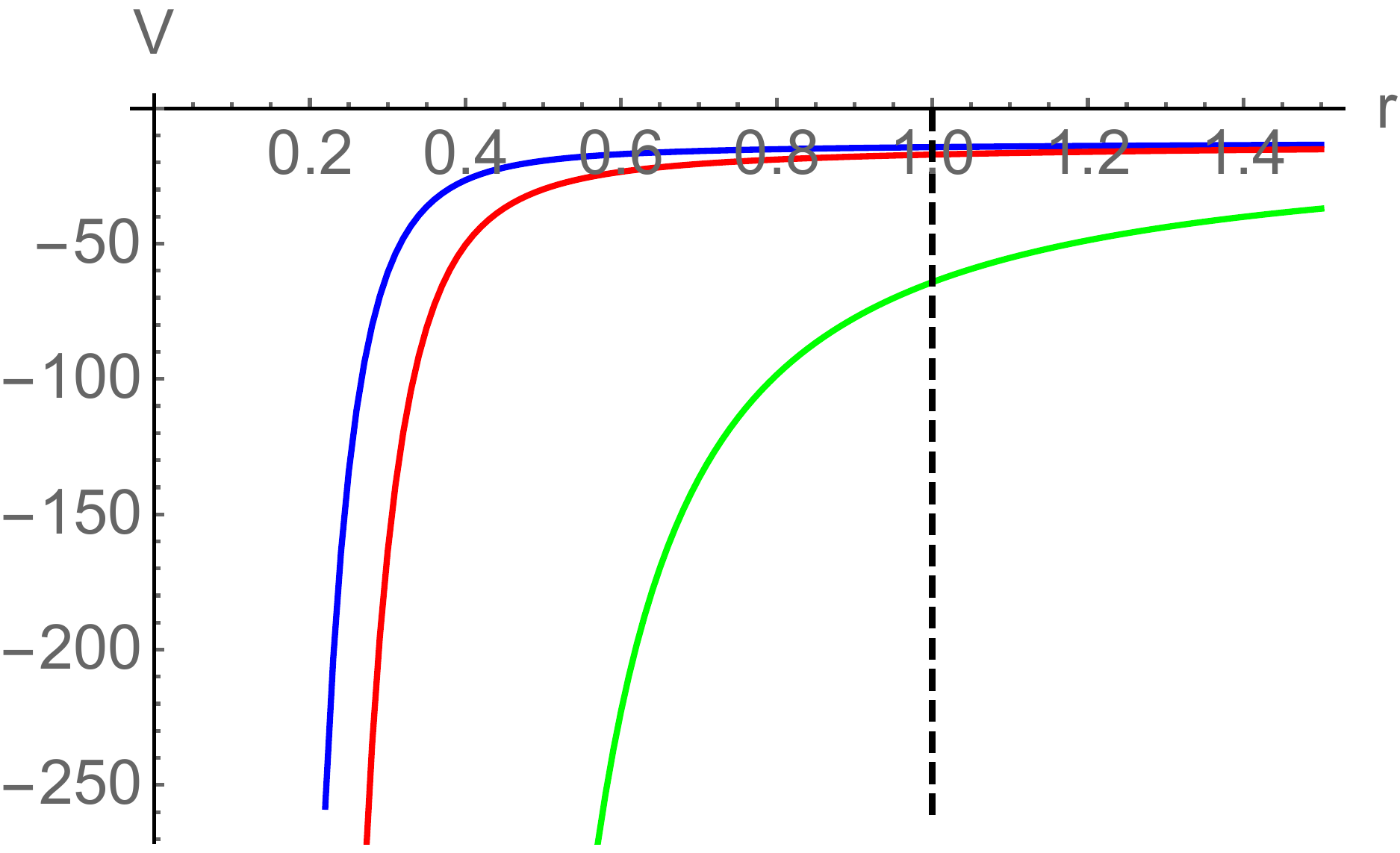}
		\includegraphics[scale=0.4]{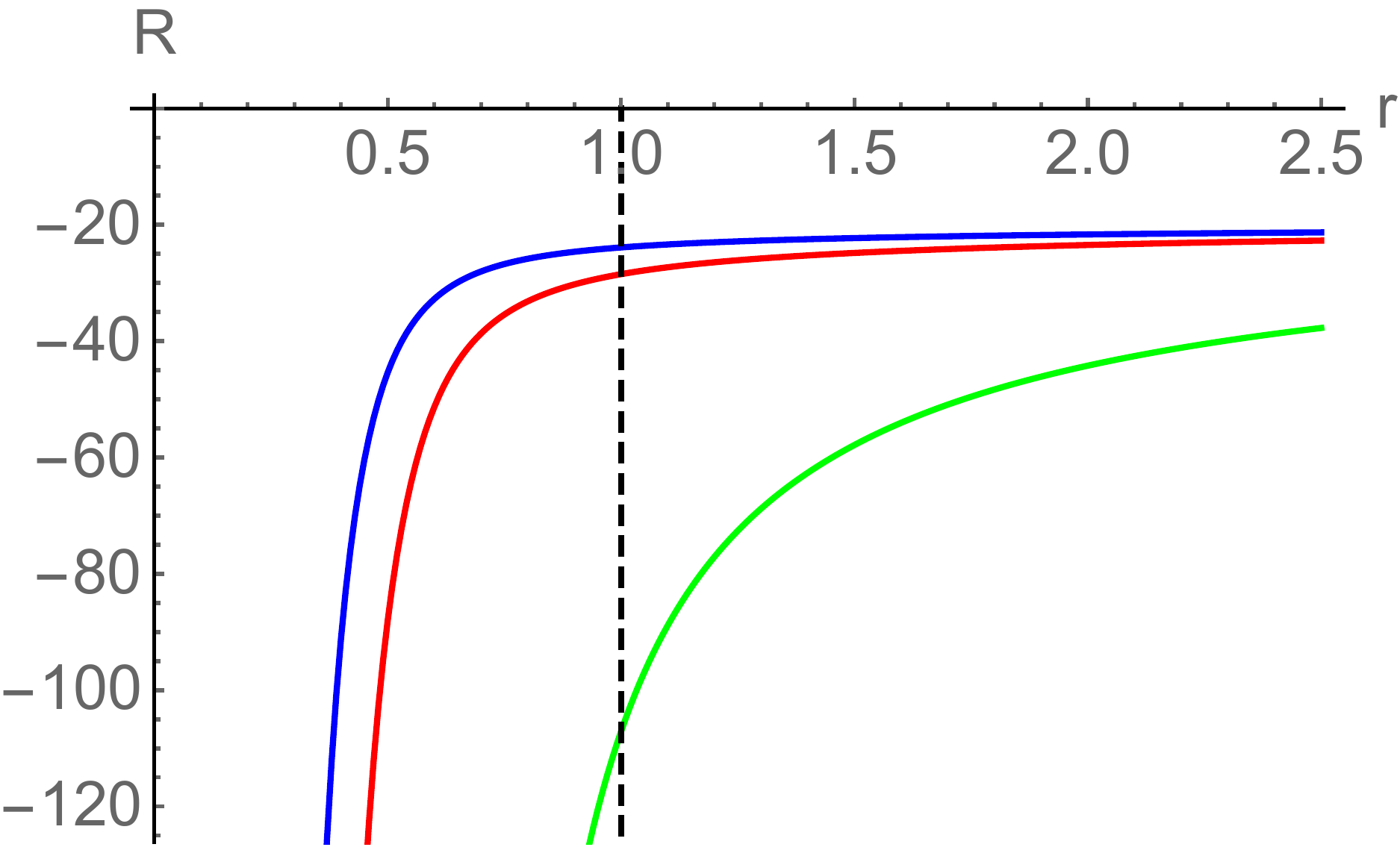}\\
		a & b
	\end{tabular}
	\caption{Panel (a) shows the variation of dilaton potential $V(r)$ for $d=4$ and $n=1$  and (b) shows the variation of Ricci scalar $R$. In both cases $r_h=1$ is considered. The blue curves indicate $a=0.05$, the red curves $a=0.1$ and the green curves $a=0.5$. The dotted line represents $r=r_h$.}
	\label{fig:rvsRiccid4n1}
\end{figure}

In the light of above discussion it is important to investigate further the inner and outer structure of the black hole. In particular, it is important to analyze whether the model under consideration contains any additional singularity or other non-trivial features inside the black hole which can potentially rule out the novel results of our model. For this purpose, in Fig. \ref{fig:rvsRiccid4n1} we have shown the behavior of dilaton potential and Ricci scalar for $d=4, n=1$. We find no evidence for any additional singularity inside the black hole. The Ricci scalar diverges only at $r=0$ and remains finite everywhere else. Moreover, the Ricci scalar approaches a constant value $R=-20$ at the asymptotic boundary, as is expected for a five dimensional asymptotic AdS space.  Similarly, the dilaton potential (as well as dilaton field itself) also does not show any non-trivial feature inside the black hole. Moreover, as was also mentioned in section 2, the potential is bounded from above by its UV boundary value \textit{i.e} $V(\infty) \geq V(r)$, satisfying the Gubser criterion to have a well defined dual boundary theory as well \cite{Gubser:2000nd}. Therefore, we can safely say that the novel features of holographic complexity in our model are genuine and not a mathematical artifact of any additional singularity. We further like to mention that these nice features of dilaton potential and Ricci scalar remain true even for other values of $d$ and $n$. Therefore, our results for the holographic complexity for other values of $d$ and $n$ in below sections will also remain true.

\subsubsection{Case 2: $d=4$, $n=2$}
\begin{figure}[htp]
	\subfigure[]{
		\includegraphics[scale=0.5]{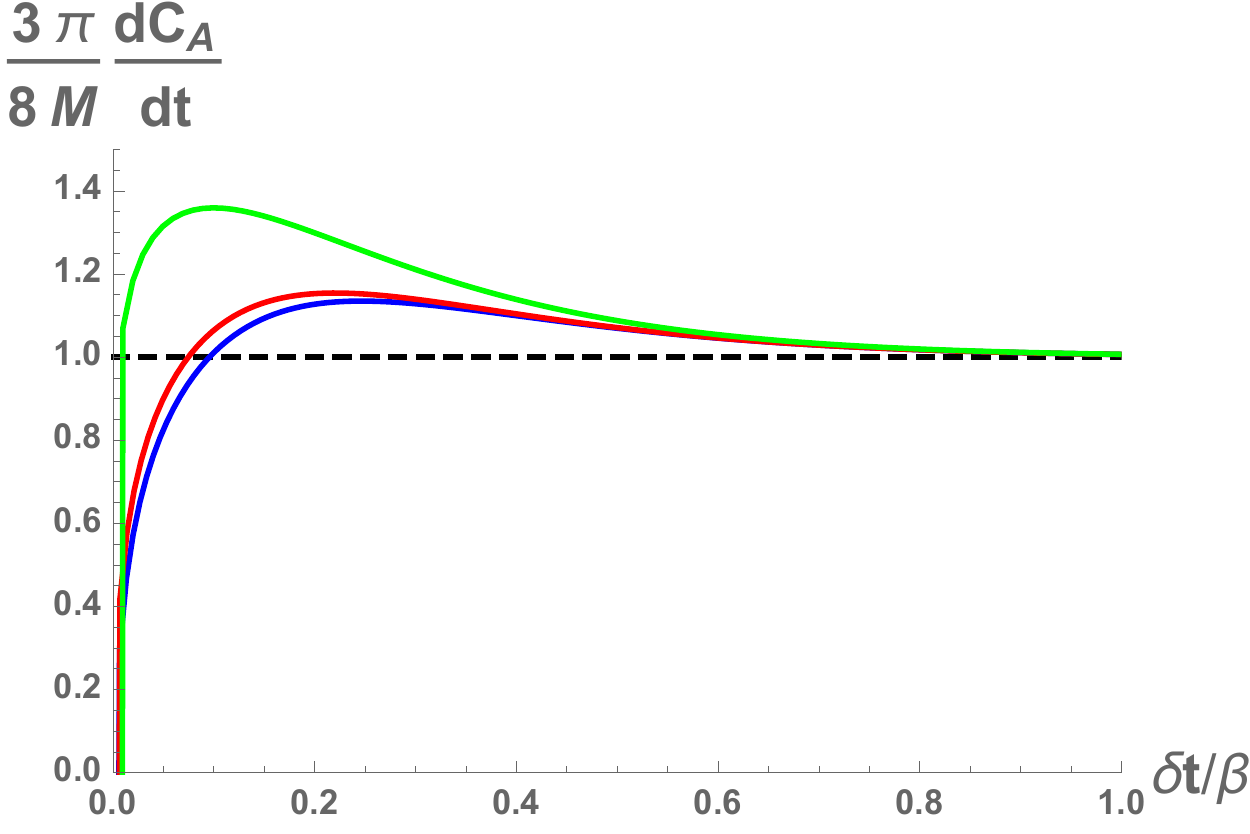}
	}
	\subfigure[]{
		\includegraphics[scale=0.5]{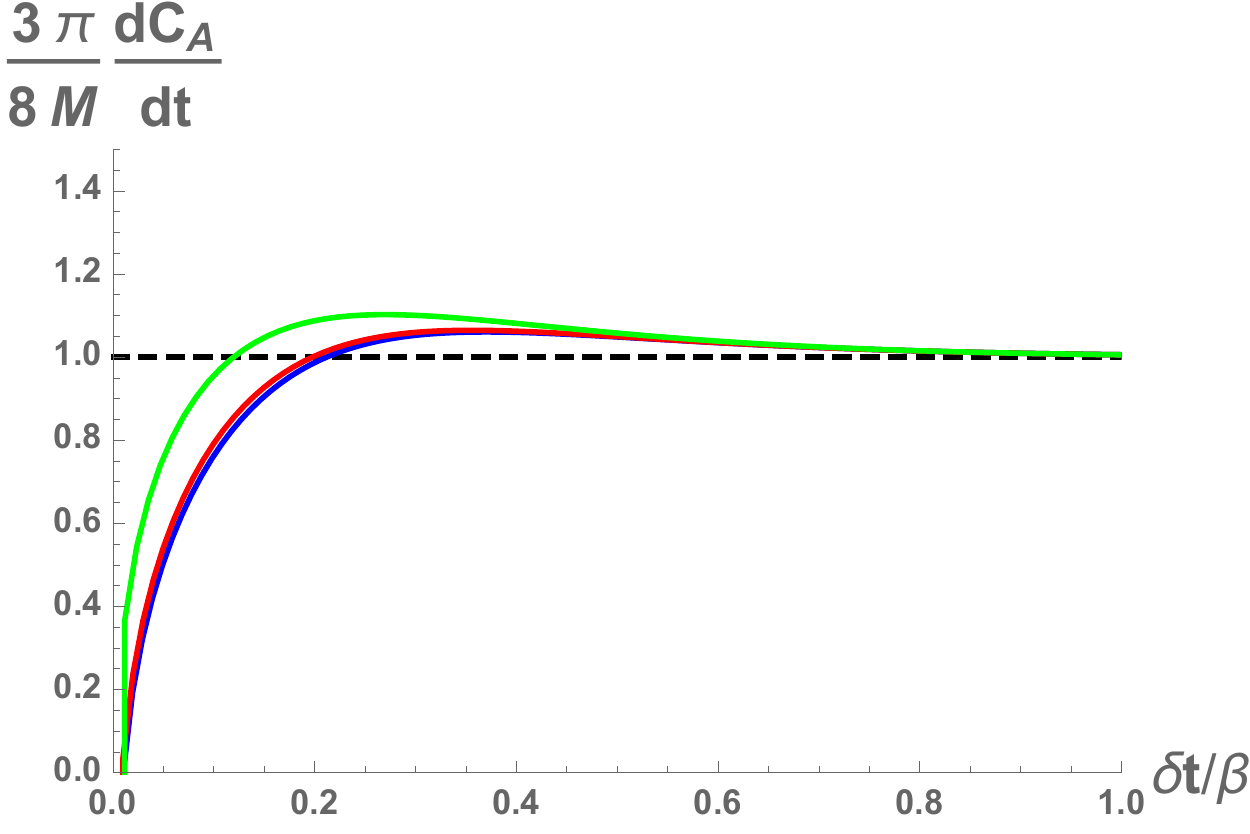}
	}
	\subfigure[]{
		\includegraphics[scale=0.5]{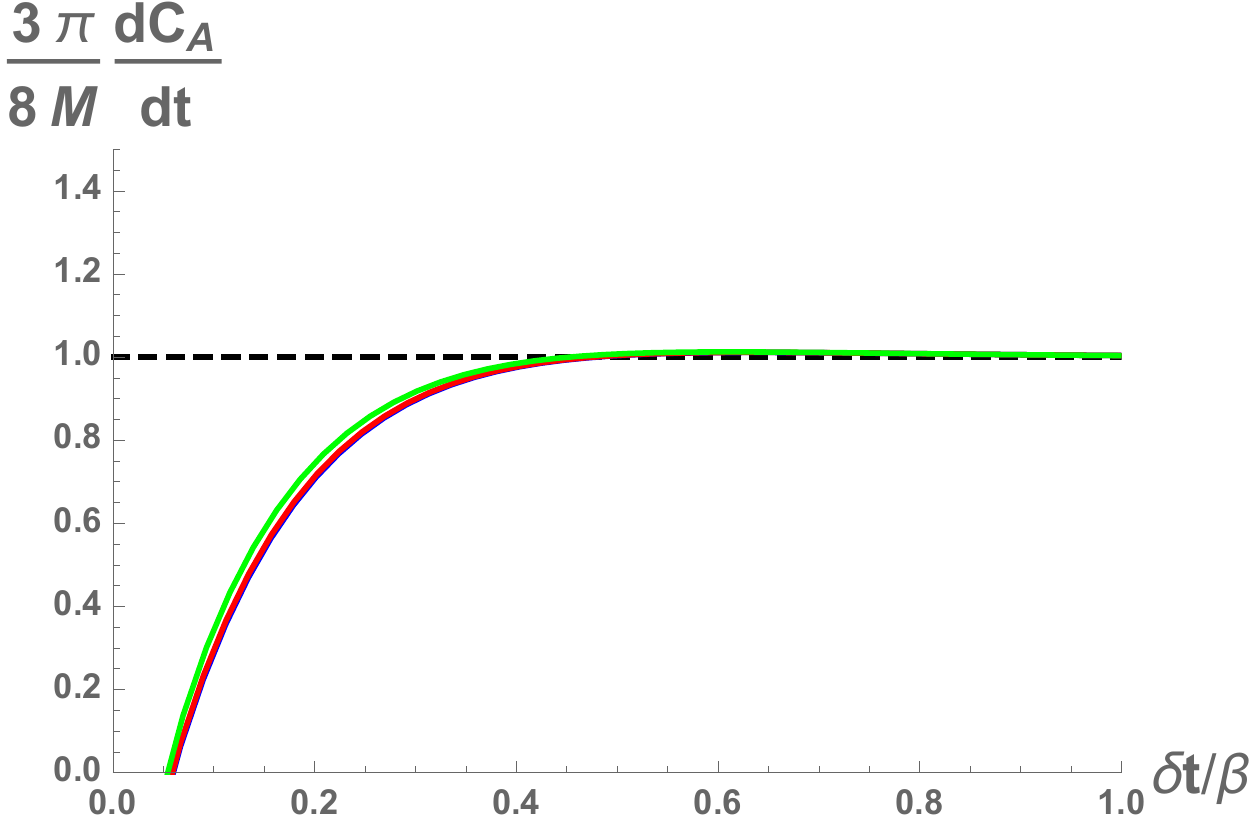}
	}
	\caption{$\frac{dC_A}{dt}$ is plotted against the dimensionless quantity $\delta t/\beta$ for $d=4$ and $n=2$. Here $\beta$ is inverse temperature. Panel (a) shows the variation of $\frac{dC_A}{dt}$ for $r_h = 1.0$, (b) for $r_h=1.5$ and for (c) for $r_h = 3.5$. In each panel blue, red and green curves correspond to $a=0.05$, $a=0.1$ and  $a=0.5$ respectively. In all cases, the dotted line represents the ratio $\frac{3 \pi}{8 M}\frac{dC_A}{dt}=1$. The values of the constants $c, \bar{c}=1$ have been set for simplicity.}
	\label{fig:dCAdtd4n2}
\end{figure}

To further support the above conclusion that the Lloyd bound is explicitly violated for the Einstein-dilaton model under consideration, we below proceed to verify whether it holds for $n=2$. For this purpose let us first calculate the late time limit of $\frac{dC_A}{dt}$. To this end, let us record the bulk, surface and joint contributions for $n=2$
\begin{eqnarray}
& & \frac{d S_{bulk}}{dt} = - \frac{V_3 }{16\pi G_5} \frac{12 a^2 r_h^2}{r_h^2 + e^{\frac{3a}{r_h^2}}\left(3a -r_h^2 \right)} \ , \\ \nonumber
& & \frac{d S_{surf}}{dt} = \frac{V_3 }{16\pi G_5} \frac{30 a^2 r_h^2}{r_h^2 + e^{\frac{3a}{r_h^2}}\left(3a -r_h^2 \right)} \ , \\ \nonumber
& & \frac{d S_{jnt}}{dt} = \frac{V_3 }{16\pi G_5} \frac{18 a^2 r_h^2}{r_h^2 + e^{\frac{3a}{r_h^2}}\left(3a -r_h^2 \right)} \,.
\end{eqnarray}
where, the following expression for $g(r)$ is used,
\begin{equation}
\label{gd4n2}
g(r)  =  1-\frac{e^{\frac{3 a}{r^2}} \left(\frac{3 a}{r^2}-1\right)+1}{e^{\frac{3
   a}{r_{h}^2}} \left(\frac{3 a}{r_{h}^2}-1\right)+1} \,.
\end{equation}
and similarly the expression for the black hole mass is,
\begin{equation}
 M = \frac{V_{3}}{32\pi G_5 } \frac{27a^2 r_h^2}{r_h^2 + e^{\frac{3a}{r_h^2}}\left(3a -r_h^2 \right)} \,.
\end{equation}
It can readily be observed that the relation between the rate of change of complexity for $n=2$ is the same as that in eq. (\ref{CAd5relation}),
\begin{equation}
\lim_{t \rightarrow \infty} \frac{dC_A}{dt} =  \frac{8M}{3\pi} \,.
\end{equation}
which again violates the Lloyd bound. We may also evaluate the general time dependence of $\frac{dC_A}{dt}$ to confirm the above late time limit by using the same method as described in the previous section. The results are shown in Fig. \ref{fig:dCAdtd4n2}

We see that Fig. \ref{fig:dCAdtd4n2} displays the same features as the $n =1$ case, thus lending further credence to our results that the Lloyd bound is violated in our Einstein-dilaton system not only at early times but at late times as well.

\subsubsection{Holographic complexity in $d=3$}

As a last verification of the conclusion that the rate of change of complexity calculated according to the CA proposal violates the Lloyd bound in our gravity model, we perform the same analysis in $AdS_4$. Since the calculations are similar to $AdS_5$, we quote the results directly for $n=1$. Let us first record the expression of $g(r)$ in this case,
\begin{equation}
g(r) = 1-\frac{\frac{e^{\frac{2 a}{r}} \left(2 a^2-2 a
   r+r^2\right)}{r^2}-1}{\frac{e^{\frac{2 a}{r_h}} \left(2 a^2-2 a
   r_h+r_{h}^2\right)}{r_{h}^2}-1} \,.
\end{equation}
The expressions for the rate of change of the bulk, surface and joint contributions in the late time limit is given by,
\begin{eqnarray}
& & \frac{d S_{bulk}}{dt} = -\frac{V_2}{16 \pi G_4}\frac{4 a^3 r_h^2}{e^{\frac{2 a}{r_h}} \left(2 a^2-2 a r_h+r_h^2\right)-r_h^2} \ , \\ \nonumber
& & \frac{d S_{surf}}{dt} = \frac{V_2}{16 \pi G_4} \frac{8 a^3 r_h^2}{e^{\frac{2 a}{r_h}} \left(2 a^2-2 a r_h+r_h^2\right)-r_h^2} \ , \\ \nonumber
& & \frac{d S_{jnt}}{dt} = \frac{V_2}{16 \pi G_4} \frac{4 a^3 r_h^2}{e^{\frac{2 a}{r_h}} \left(2 a^2-2 a r_h+r_h^2\right)-r_h^2} \,.
\end{eqnarray}
For this configuration, ($n=1$ and $d=3$), the expression for the mass of the black hole is given by,
\begin{equation}
M = \frac{V_2}{6 \pi G_4} \frac{a^3 r_h^2}{ \left[e^{\frac{2 a}{r_h}} \left(2 a^2-2 a r_h+r_h^2\right)-r_h^2 \right]} \,.
\end{equation}
From the above equations, we may infer the following relation between $dC_A/dt$ and $M$,
\begin{equation}
\lim_{t \rightarrow \infty} \frac{dC_A}{dt} = \frac{3M}{\pi} \,.
\end{equation}

\begin{figure}[htp]
	\subfigure[]{
		\includegraphics[scale=0.5]{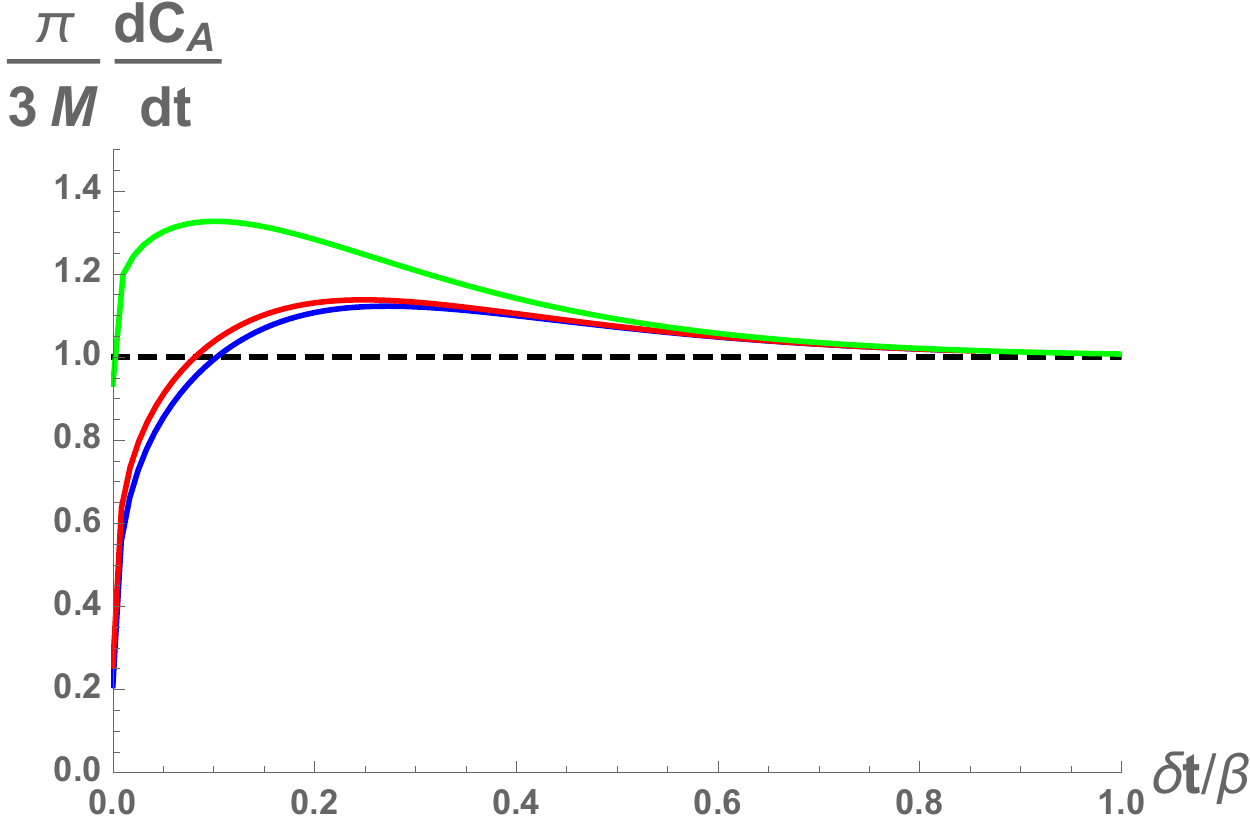}
	}
	\subfigure[]{
		\includegraphics[scale=0.5]{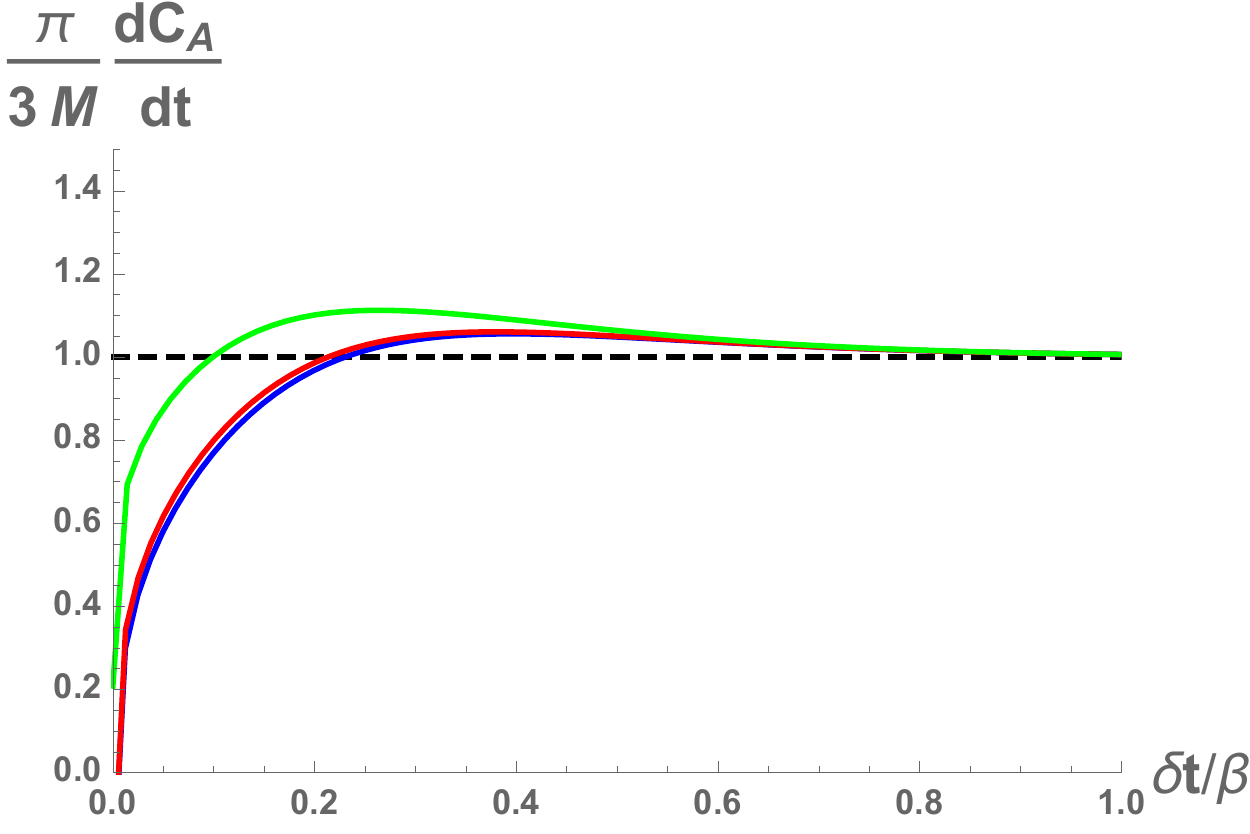}
	}
	\subfigure[]{
		\includegraphics[scale=0.5]{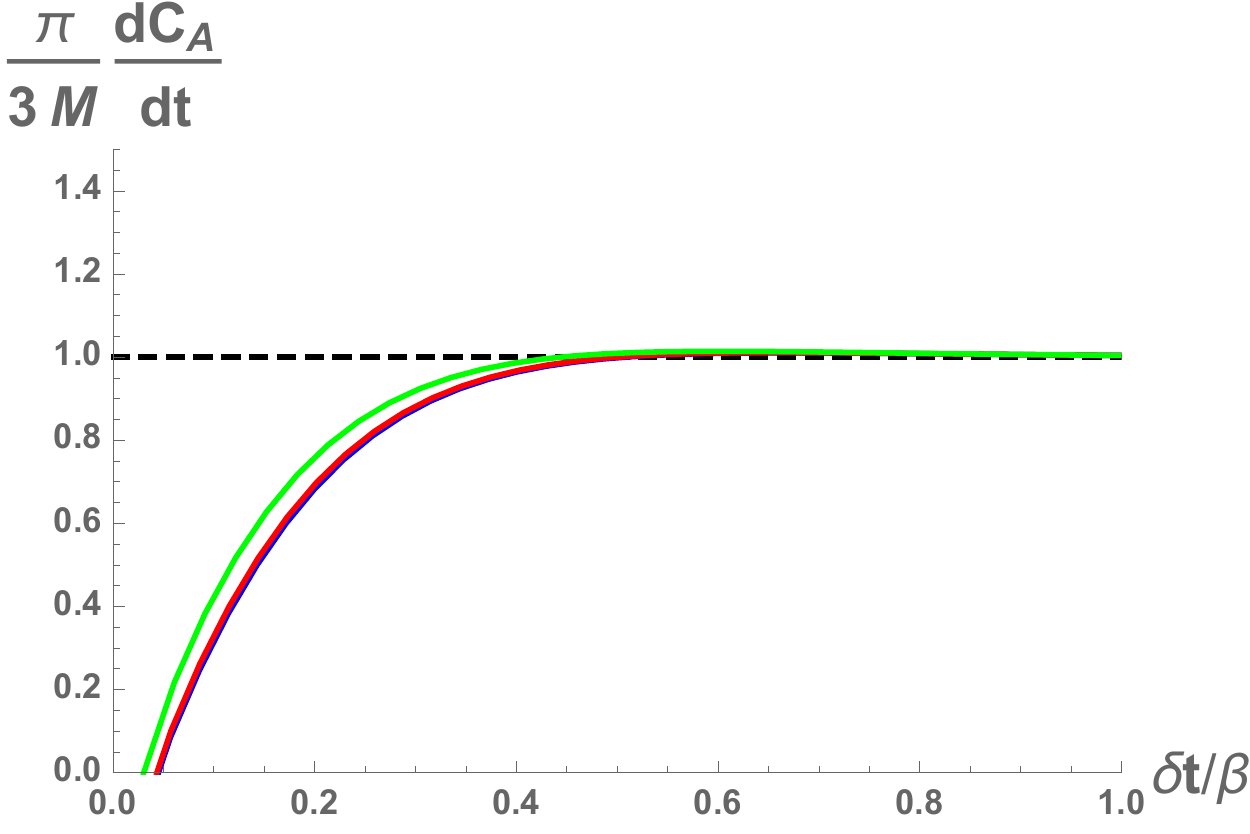}
	}
	\caption{$\frac{dC_A}{dt}$ is plotted against the dimensionless quantity $\delta t/\beta$ for $d=3$ and $n=1$. here $\beta$ is inverse temperature. Panel (a) shows the variation of $\frac{dC_A}{dt}$ for $r_h = 1.0$, (b) for $r_h=1.5$ and for (c) for $r_h = 3.5$. In each panel blue, red and green curves correspond to $a=0.05$, $a=0.1$ and  $a=0.5$ respectively. In all cases, the dotted line represents the ratio $\frac{\pi}{3 M}\frac{dC_A}{dt}=1$. The values of the constants $c, \bar{c}=1$ have been set for simplicity.}
	\label{fig:dCAdtd3n1}
\end{figure}

The same result (which is not recorded here for brevity) may be shown to hold for $n=2$ in $AdS_4$. The complete time dependence of $\frac{dC_A}{dt}$ is shown in Fig. \ref{fig:dCAdtd3n1}, where we see quantitatively the same features as were found in $d=4$ case.\\

In summary to the results of this section, we may state that for the Einstein-dilaton model considered in this paper, the Lloyd bound is violated generically. While the time-rate of change of complexity is proportional to the mass of the black hole, the (dimension-dependent) proportionality constant ensures that Lloyd bound is violated and the rate of increase of complexity with time actually exceeds $2M/\pi$. We have further summarized our results for the late time limit of $dC_A/dt$ in various dimensions in Table \ref{dcadtvsdtable}, asserting that the Lloyd bound is indeed violated in all spacetime dimensions.
{ \setlength\extrarowheight{8pt}
\begin{table}
\begin{center}
\begin{tabular}{ | m{1.2cm} | m{3.4cm}| m{3.4cm} |}
\hline
  & \hspace{1cm} $n=1$ & \hspace{1cm}  $n=2$  \\
\hline
$d=3$ & $\underset{t \to \infty}{\lim} \frac{dC_A}{dt}=3 M/\pi$  &  $\underset{t \to \infty}{\lim} \frac{dC_A}{dt}=3 M/\pi$ \\
\hline
$d=4$ & $\underset{t \to \infty}{\lim} \frac{dC_A}{dt}=8 M/ 3\pi$ &  $\underset{t \to \infty}{\lim} \frac{dC_A}{dt}=8 M/ 3\pi$ \\
\hline
$d=5$ & $\underset{t \to \infty}{\lim} \frac{dC_A}{dt}=5M/2\pi$ & $\underset{t \to \infty}{\lim} \frac{dC_A}{dt}=5M/2\pi$ \\
\hline
$d=6$ & $\underset{t \to \infty}{\lim} \frac{dC_A}{dt}=12M/5\pi$ & $\underset{t \to \infty}{\lim} \frac{dC_A}{dt}=12M/5\pi$ \\
\hline
$d=7$ & $\underset{t \to \infty}{\lim} \frac{dC_A}{dt}=7M/3\pi$ & $\underset{t \to \infty}{\lim} \frac{dC_A}{dt}=7M/3\pi$  \\
\hline
\end{tabular}
\end{center}
\caption{The late time limit of $dC_A/dt$ in various dimensions.}
\label{dcadtvsdtable}
\end{table}
}

\section{Complexity using CV proposal}
\label{CVsection}

In this section, we evaluate the complexity using the CV proposal. This has been done for Schwarzschild and Reissner-Nordstrom black holes in various dimensions and horizon topologies in \cite{Carmi:2017jqz}. For our Einstein-dilaton system, we proceed to calculate the complexity following a similar procedure. Our main aim in this section is to derive the complexity according to the CV proposal for comparison with the CA result. As in the previous section, we shall concentrate on $d=4$ and record the expressions for two different functional forms of the scale factor $A(r)=-a/r^n$, namely for $n=1, 2$. At the end of the section, we shall also present the results for $AdS_4$ for completeness and comparison.

\begin{figure}[h]
	\includegraphics*[width=3.5in,height=2.8in]{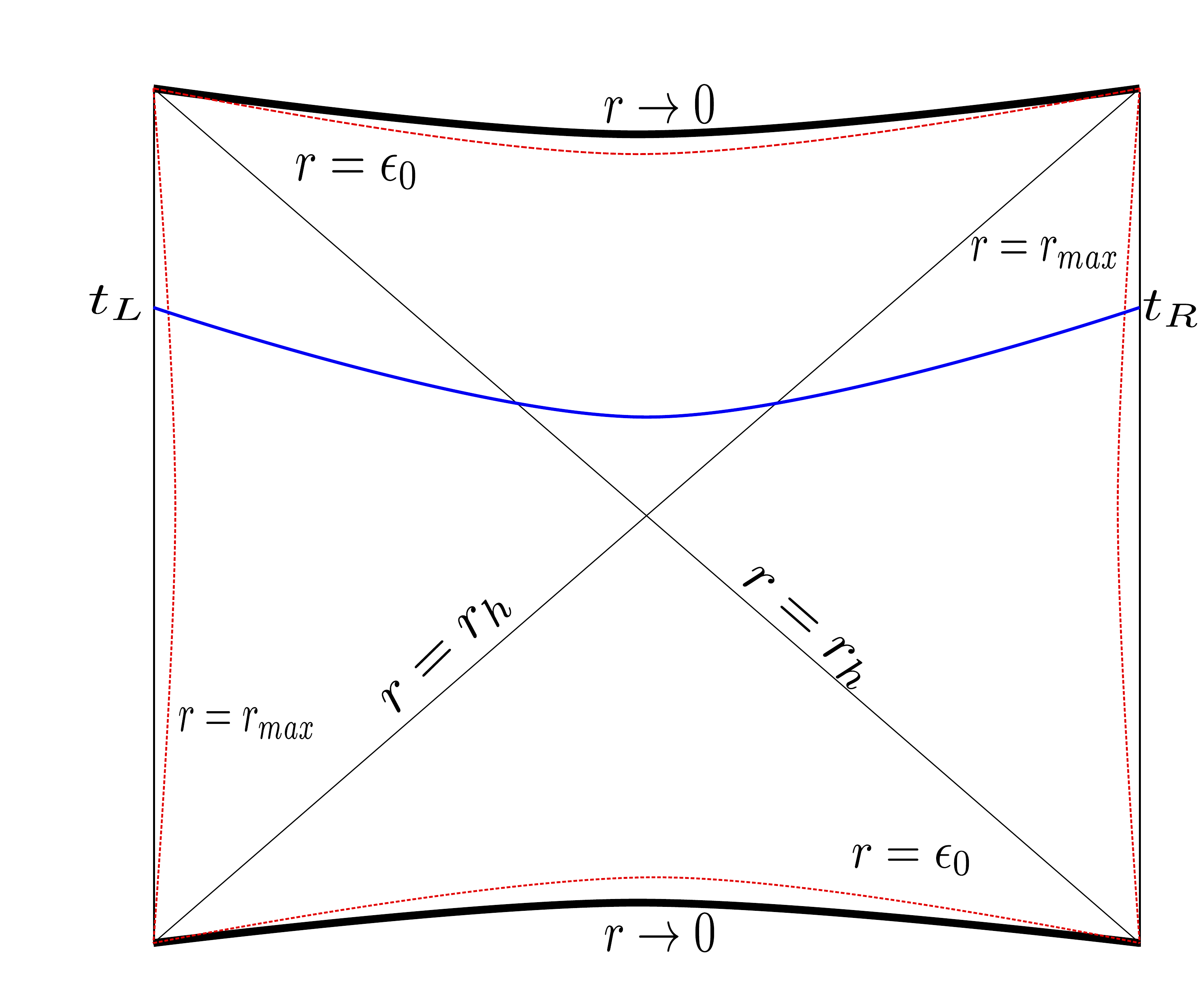}
	\caption{Representation of maximal volume surface connecting the two sides of the eternal $AdS$ black hole. The CV conjecture implies that the complexity is proportional to the maximal volume.}
	\label{fig:dCVdt}
\end{figure}

It is convenient to calculate the complexity using the metric given in $(v,r)$ coordinates in eq. (\ref{Eddington-Finkelstein}). According to the CV proposal, the rate of change of complexity is given by the volume of the co-dimension one extremal surface ending at constant time in the two asymptotic boundaries (see Fig. \ref{fig:dCVdt}),
\begin{equation}
C_V = \frac{\text{max}(\mathcal{V})}{G_{d+1} \ell} \,.
\end{equation}
We assume that the co-dimension one surface described above has the same symmetry of the horizon, i.e. the surface has no functional dependence on the coordinates $y_i$. In order to calculate $C_V$, we further parameterized the extremal surface  as follows,
\begin{equation}
r \equiv r(\lambda), ~ v \equiv v(\lambda) \,.
\end{equation}
With this parametrization the induced line element reduces to,
\begin{equation}
ds_{\Sigma}^2 = \left( -G(r) \dot{v}^2 + 2 e^{2A(r)} \dot{v} \dot{r} \right)d\lambda^2 + r^2 e^{2A(r)} d\vec{y}_{d-1}^2 \,.
\end{equation}
The maximal volume is then obtained by extremizing the following expression,
\begin{eqnarray}
\label{volume}
\mathcal{V} &=& V_{d-1} \int d\lambda \ r^{d-1} e^{(d-1) A(r)} \sqrt{-G(r)\dot{v}^2 + 2 e^{2A(r)}\dot{v} \dot{r}} \ , \\ \nonumber
&=& V_{d-1} \int d\lambda \ \mathcal{L}(\dot{v}, \dot{r}, r) \,.
\end{eqnarray}
Since the ``Lagrangian ($\mathcal{L}$)'' does not depend on $v$ we have a conserved quantity which is given by,
\begin{eqnarray}
\label{conservedenergy}
E &=& -\frac{\partial \mathcal{L}}{\partial \dot{v}} \ , \\ \nonumber
&=& r^{d-1} e^{(d-1)A(r)} \frac{G(r)\dot{v}-e^{2A(r)}\dot{r}}{\sqrt{-G(r)\dot{v}^2 + 2 e^{2A{r}}\dot{v} \dot{r}}} \,.
\end{eqnarray}
Also, using the reparametrization invariance of eq. (\ref{volume}) we may fixed the radial volume element, \textit{i.e}
\begin{equation}
\label{reparam}
r^{d-1} e^{(d-1)A(r)} \sqrt{-G(r)\dot{v}^2 + 2 e^{2A(r)}\dot{v} \dot{r}} = 1 \,.
\end{equation}
Using eqs. (\ref{conservedenergy}) and (\ref{reparam}) we get the following equations for $r(\lambda)$ and $v(\lambda)$,
\begin{eqnarray}
\label{mastereqn}
E &=& r^{2(d-1)} e^{2(d-1)A{r}} \left(G(r)\dot{v} - e^{2A(r)}\dot{r} \right) \ , \\ \nonumber
\dot{r}^2 r^{2(d-1)} e^{2(d+1)A(r)} &=& G(r) + E^2 r^{-2(d-1)} e^{-2(d-1)A(r)} \,.
\end{eqnarray}
The extremal volume is then reduced to,
\begin{eqnarray}
\mathcal{V} &=& 2V_{d-1} \int \frac{dr}{\dot{r}} \ , \\ \nonumber
&=& 2V_{d-1} \int_{r_{min}}^{\infty} dr \frac{r^{2(d-1)} e^{2 d A(r)}}{\sqrt{G(r)r^{2(d-1)} e^{2(d-1)A(r)}+E^2}} \,.
\end{eqnarray}
where the turning point $r_{min}$ of the extremal surface, which lies inside the horizon, in the above equation is obtained by setting $\dot{r}|_{r=r_{min}} = 0$ which yields,
\begin{equation}
\label{rmin}
G(r_{min}) r_{min}^{2(d-1)} e^{2(d-1)A(r_{min})} + E^2 = 0 \,.
\end{equation}
Taking into account that $r_{min} < r_h$ and consequently $G(r_{min})<0$, we may conclude that $E<0$. So we have,
\begin{equation}
E = -\sqrt{|G(r_{min})|}r_{min}^{d-1} e^{(d-1)A(r_{min})} \,.
\end{equation}

Now using eq. (\ref{mastereqn}), it is also straightforward to derive the following equation,
\begin{equation}
\label{similarintegral}
t_R + r^{*}_{\infty} - r^{*}(r_{min}) = \int_{r_{min}}^{\infty} dr e^{2A(r)} \left[\frac{E}{G(r)\sqrt{G(r)r^{2(d-1)} e^{2(d-1)A(r)}+E^2}} + \frac{1}{G(r)}\right] \,.
\end{equation}
where, using the symmetry of our configuration, we have set $t=0$ at $r=r_{min}$.  After some algebraic manipulation, eq. (\ref{similarintegral}) can further be rewritten as
\begin{equation}
\frac{\mathcal{V}}{2V_{d-1}} = \int_{r_{min}}^{\infty} e^{2A(r)} \left[\frac{\sqrt{E^2+G(r)r^{2(d-1)} e^{2(d-1)A(r)}}}{G(r)} + \frac{E}{G(r)} \right] - E(t_R + r^{*}_{\infty}-r^{*}(r_{min}))\,.
\end{equation}
Taking derivative of the above expression with respect to $t_R$ we get,
\begin{eqnarray}
\frac{1}{2V_{d-1}} \frac{d\mathcal{V}}{dt_R} = \int_{r_{min}}^{\infty} e^{2A(r)} \left[\frac{E}{G(r)\sqrt{E^2 + G(r)r^{2(d-1)} e^{2(d-1)A(r)}}} + \frac{1}{G(r)}\right]  \frac{dE}{dt_R} \\ \nonumber
 - \frac{d E}{dt_R} (t_R+r^{*}_{\infty}-r^{*}(r_{min})) - E \,.
\end{eqnarray}
Now, since $\frac{dE}{dt_R}$ is constant for a particular extremal surface it may be taken outside the integral. Further, noting the similarity of the first term with the R.H.S of eq. (\ref{similarintegral}), we can reduce the above expression to
\begin{equation}
\frac{d\mathcal{V}}{dt_R} = -2V_{d-1} E \,.
\end{equation}
Since we are considering a symmetric configuration $t_L = t_R = \frac{t}{2}$, the rate of change of complexity is therefore given by,
\begin{equation}
\label{CV}
\frac{dC_V}{dt} = \frac{V_{d-1}}{G_{d+1}} \sqrt{-G(r_{min})}r_{min}^{2(d-1)} e^{2(d-1)A(r_{min})} \,.
\end{equation}
Let us first discuss the late time limit of $dC_V/dt$. The late time limit of $dC_V/dt$ can be obtained by noting that the concerned maximal surface will be approximately tangent to a special
constant $r=\tilde{r}_{min}$ surface inside the black hole horizon. Recalling that the minimum radius $r_{min}$ is given by eq. (\ref{rmin}), we may conclude by arguments similar to \cite{Carmi:2017jqz} that the value of $r_{min}$ at late times is given by the extremum of the function $W(r_{min}) \equiv \sqrt{-G(r_{min})}r_{min}^{d-1} e^{(d-1)A(r_{min})}$,
\begin{equation}
W^{\prime} (\tilde{r}_{min})= 2 |G(\tilde{r}_{min})| +2 \tilde{r}_{min} |G(\tilde{r}_{min})| A^{\prime}(\tilde{r}_{min}) + |G^{\prime}(\tilde{r}_{min})| = 0 \,.
\label{rtilmin}
\end{equation}
where, $\tilde{r}_{min}$ is both a root of eq. (\ref{rmin}) as well as extremum of $W(r)$. Therefore, in the late time limit we have
\begin{equation}
\label{inequality1}
\lim_{t\to\infty} \frac{dC_V}{dt} = \frac{V_{d-1}}{G_{d+1}} \sqrt{-G(\tilde{r}_{min})}\tilde{r}_{min}^{d-1} e^{(d-1)A(\tilde{r}_{min})} \,.
\end{equation}
Therefore, once we have the solution for $\tilde{r}_{min}$ by solving eq. (\ref{rtilmin}), we can subsequently obtain $dC_V/dt$ in the late time limit by evaluating eq. (\ref{inequality1}). Unfortunately, eq. (\ref{rtilmin}) cannot be solved analytically for any non-zero $a$ and hence we do not have analytic results for $\frac{dC_V}{dt}$ in our model. However, it can be easily solved numerically. In the subsequent subsections, we will study the variation of $\lim_{t\to\infty}\frac{dC_V}{dt}$ against $r_h$ for the same specific cases which were studied for the CA proposal in section \ref{CAsection} of this paper.

In addition to the late time limit, we may also plot the rate of change of complexity with time to investigate the approach to its late-time value. For this, we utilize eq. (\ref{similarintegral}), whence we obtain the expression for the boundary time $t$,
\begin{equation}
\label{boundarytime}
t = 2 \int_{r_{min}}^{\infty} dr e^{2A(r)} \left[\frac{E}{G(r)\sqrt{G(r)r^{2(d-1)} e^{2(d-1)A(r)}+E^2}} \right] \, .
\end{equation}
For numerical purpose, it is more convenient to use the dimensionless variables $s = r/r_h$. Using $s$, the integral in eq. (\ref{boundarytime}) can be evaluated numerically for any values of $d$ and $n$. In the subsequent sections, the horizon radius is measured in units of AdS length scale $L$ (which, as mentioned earlier, has been set to unity).

\subsection{Case 1: $d=4$, $n=1,2$}

With the expressions for $\tilde{r}_{min}$ determined from eq. (\ref{rtilmin}), it is straightforward to substitute it in eq. (\ref{CV}) to find an expression for the complexity as a function of the horizon radius $r_h$ and time $t$. Before presenting the respective plots, a word about the late time limit of the rate of change of complexity is in order here. As was deduced in \cite{Stanford:2014jda}, the high temperature limit of the rate of change of complexity for $AdS$-Schwarzschild black hole with planar horizon is given by,
\begin{equation}
\frac{dC_V}{dt} = \frac{8\pi M}{d-1}
\end{equation}
For the present set-up, it is difficult to proceed analytically to confirm the above limit. However, as we will see shortly using numerical calculations,  the time rate of change of complexity does indeed asymptote to the above value for large radii in our model as well.

\begin{figure}
	\begin{tabular}{cc}
		\includegraphics[scale=0.5]{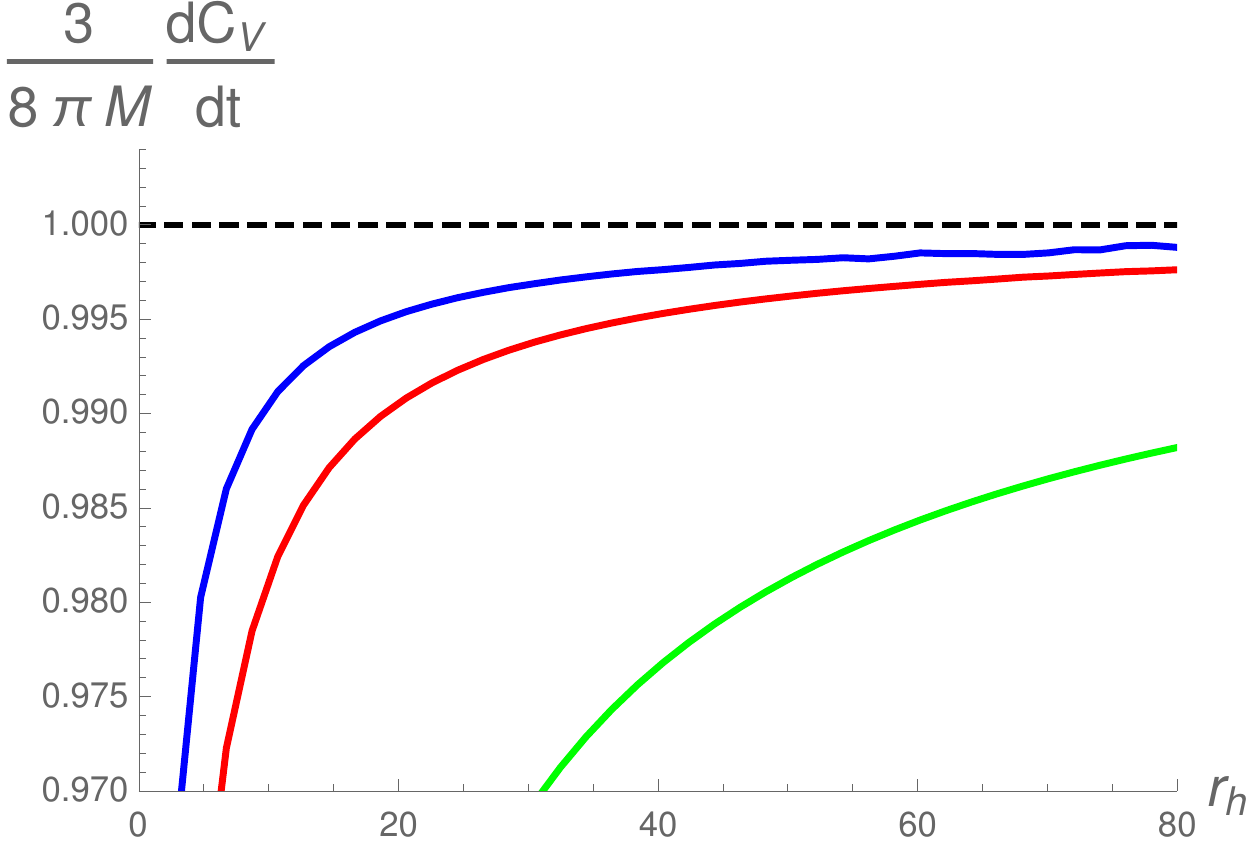}
		\includegraphics[scale=0.5]{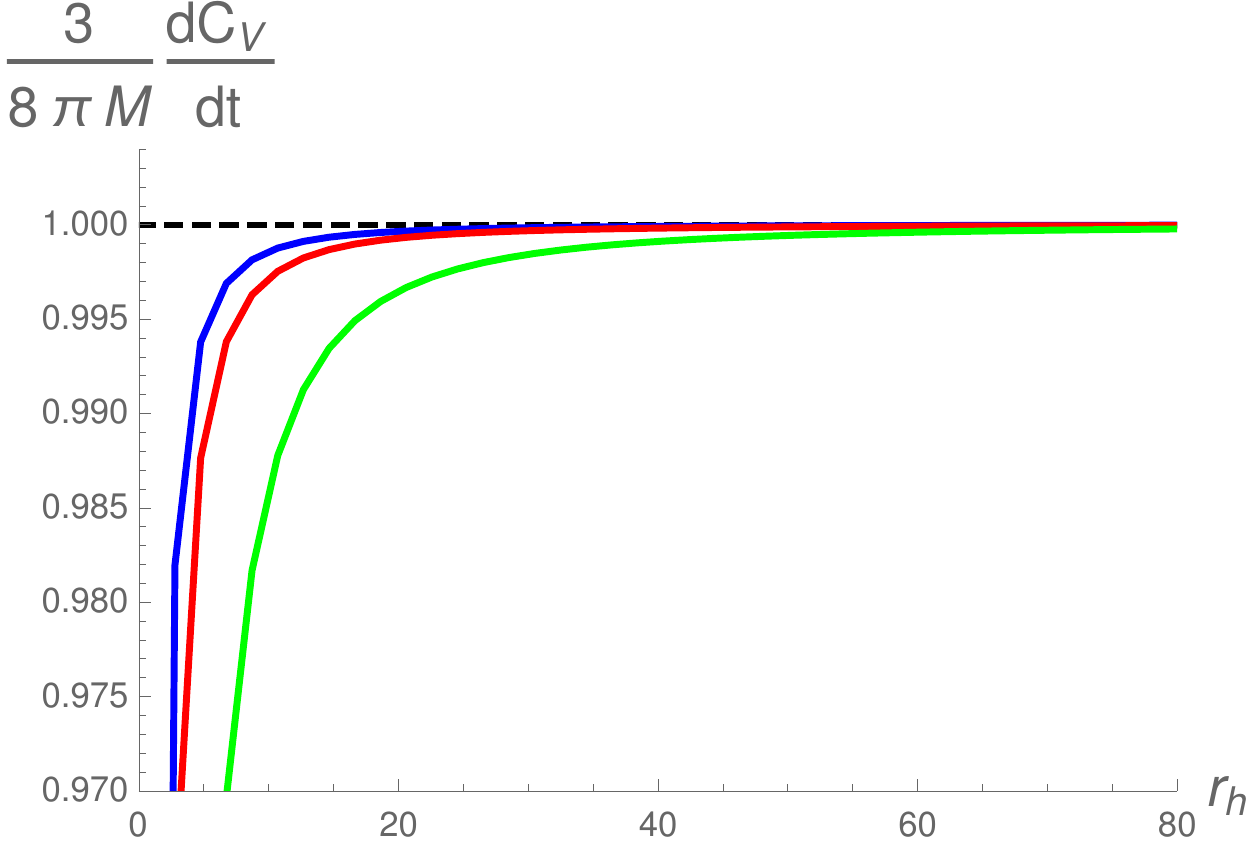}\\
		a & b
	\end{tabular}
	\caption{Panel (a) shows the variation of $\frac{dC_V}{dt}$ vs $r_h$ for $d=4$ and $n=1$ in the late time limit  and (b) shows the same variation for $d=4$ and $n=2$. In both cases, the dotted line represents the ratio $\frac{3}{8\pi M}\frac{dC_V}{dt}$. The blue curves indicate $a=0.05$, the red curves $a=0.1$ and the green curves $a=0.5$.}
	\label{fig:dCVdtd4}
\end{figure}

The numerical results for $\frac{dC_V}{dt}$ in the late time limit are presented in Fig. (\ref{fig:dCVdtd4}). Here, we have again used $n=1, 2$ with different values of the parameter $a$. It is interesting to note that, unlike the complexity=action case, the rate of change of complexity with volume conjecture matches the results obtained in \cite{Stanford:2014jda}. In particular, it may be observed from Fig. \ref{fig:dCVdtd4} that
\begin{equation}
\label{inequality}
\lim_{t\to\infty} \frac{dC_V}{dt} \leq \frac{8\pi M}{d-1} \,.
\end{equation}
which is same as found in \cite{Stanford:2014jda}. We can see that the asymptotic (maximum) value that is attained by the complexity increases to unity with increase of horizon radius $r_h$, irrespective of the value of $a$. Moreover, $dC_V/dt$ approaches this late time limit from below. This result can be traced back to the fact that $W''(\tilde{r}_{min})$ is negative due to $E<0$.

We find a few differences from the $AdS$-Schwarzschild case as well. In particular, in \cite{Carmi:2017jqz} it was observed that the above inequality was always saturated for the planar black holes for all $r_h$, while for spherical black holes, the inequality was saturated only at higher values of $r_h$ i.e. at higher $T$. However, in our case, even for the planar black hole, the above inequality gets saturated only at higher $T$. Moreover, we further find that for larger values of $a$, the temperature at which $\frac{dC_V}{dt}$ saturates is also increased. This behaviour can be clearly seen from Fig. \ref{fig:dCVdtd4} by comparing blue, red and green curves. Another observation is that for higher $n$, $dC_A/dt$ reaches its asymptotic value at lower temperatures.

\begin{figure}
\subfigure[]{
\includegraphics[scale=0.5]{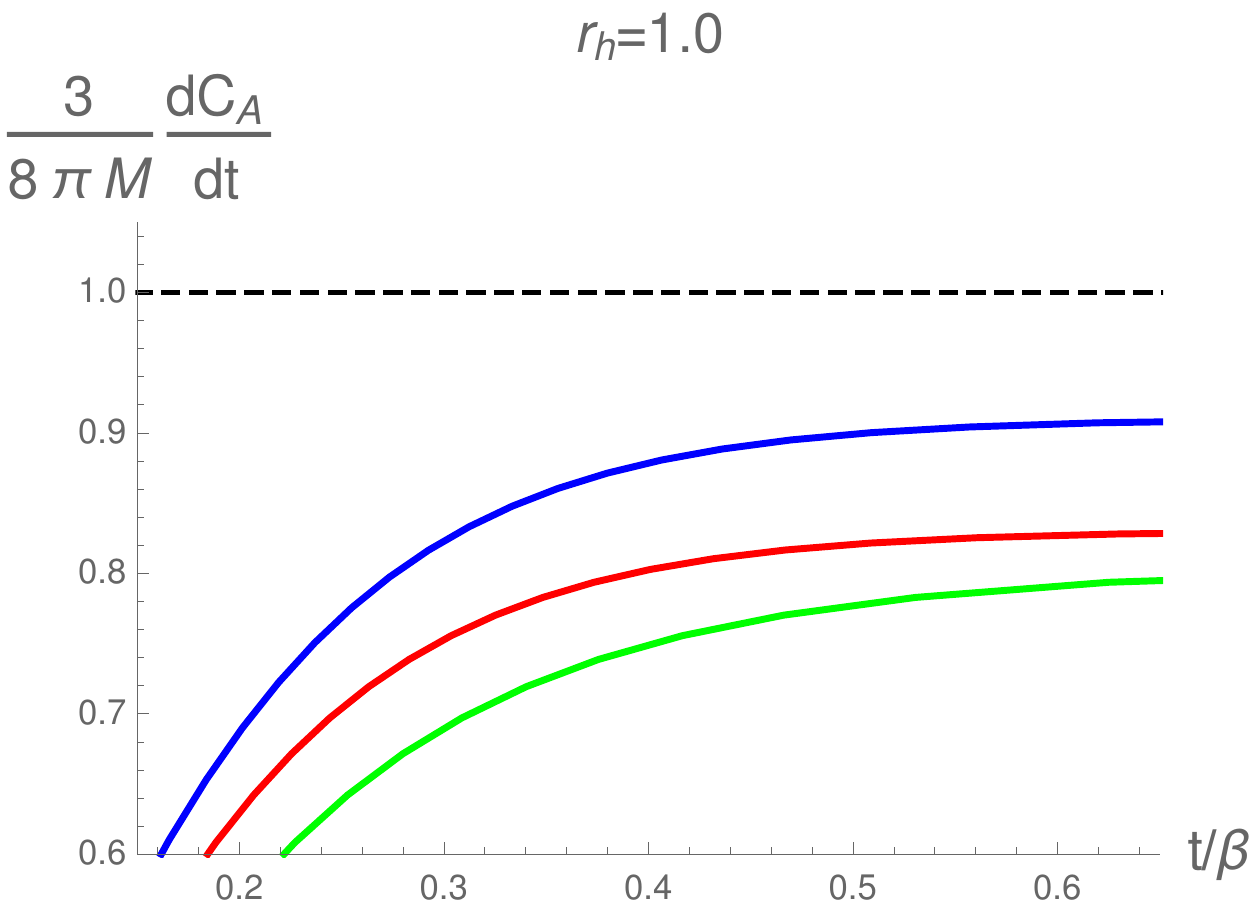}
}
\subfigure[]{
\includegraphics[scale=0.5]{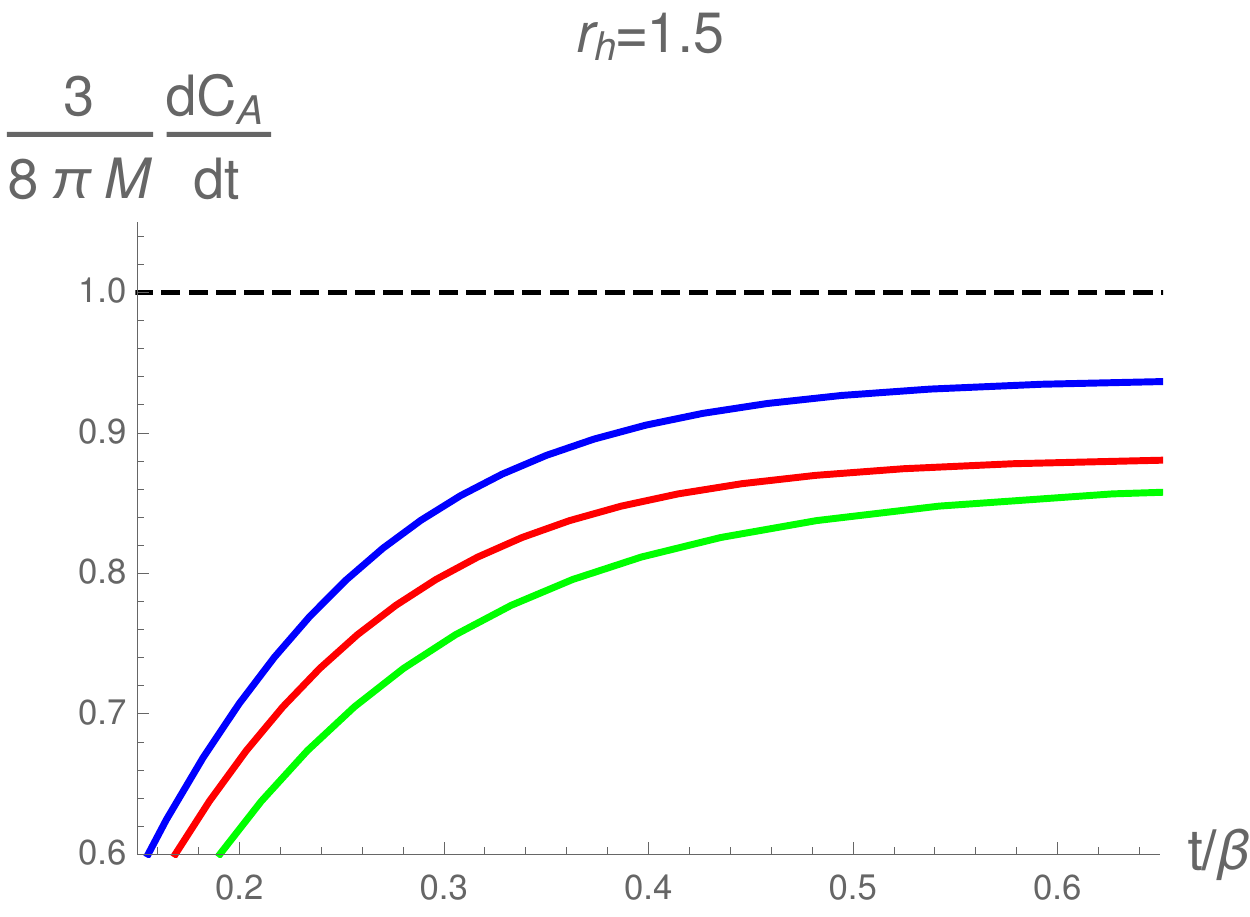}
}
\subfigure[]{
\includegraphics[scale=0.5]{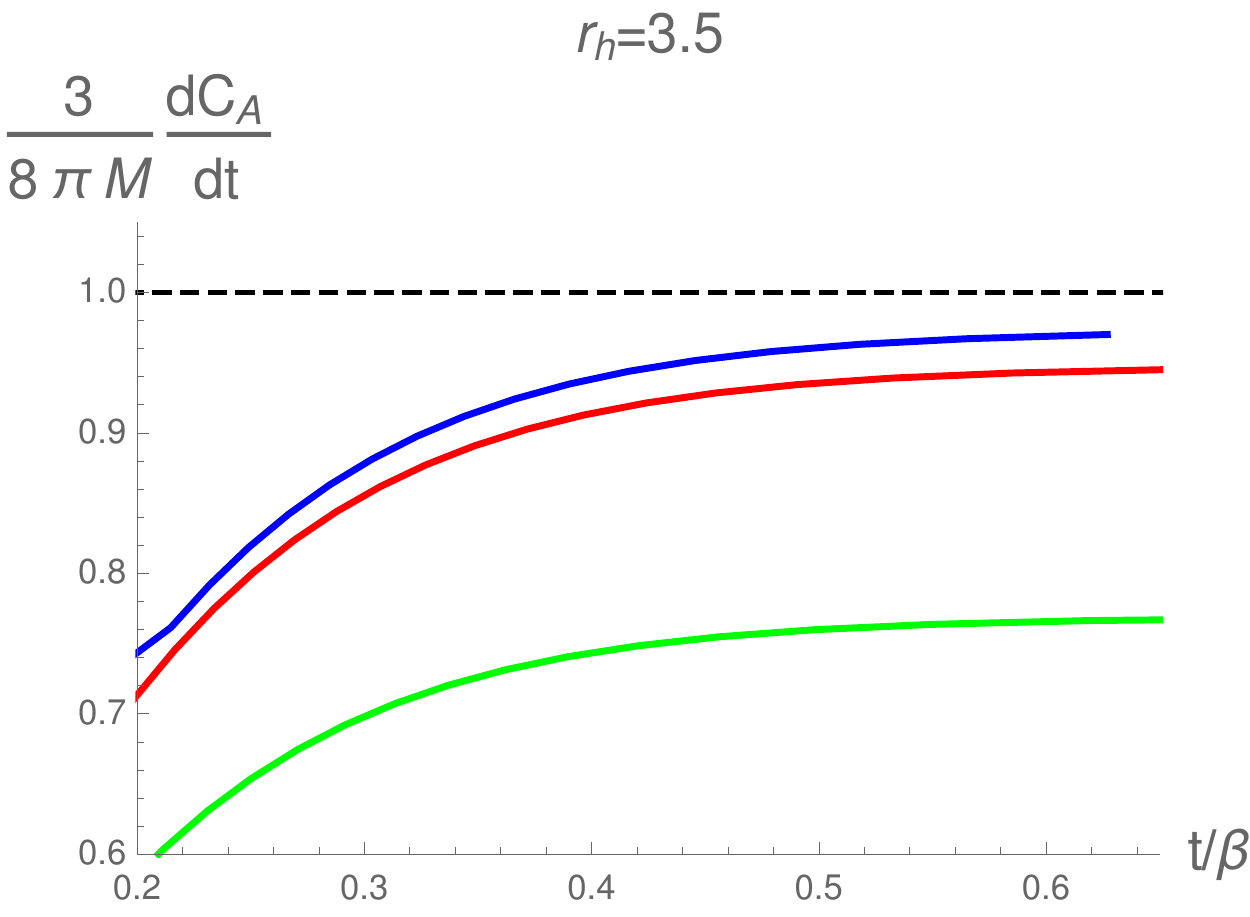}
}
\caption{$\frac{dC_V}{dt}$ as a function of $t/\beta$ for different values of the horizon radius $r_h$. Here we have used $d=4$ and $n=1$, and as before the blue curves indicate $a=0.05$, the red curves $a=0.1$ and the green curves $a=0.5$.}
\label{fig:dCVdtd4n1}
\end{figure}

\begin{figure}
	\subfigure[]{
		\includegraphics[scale=0.5]{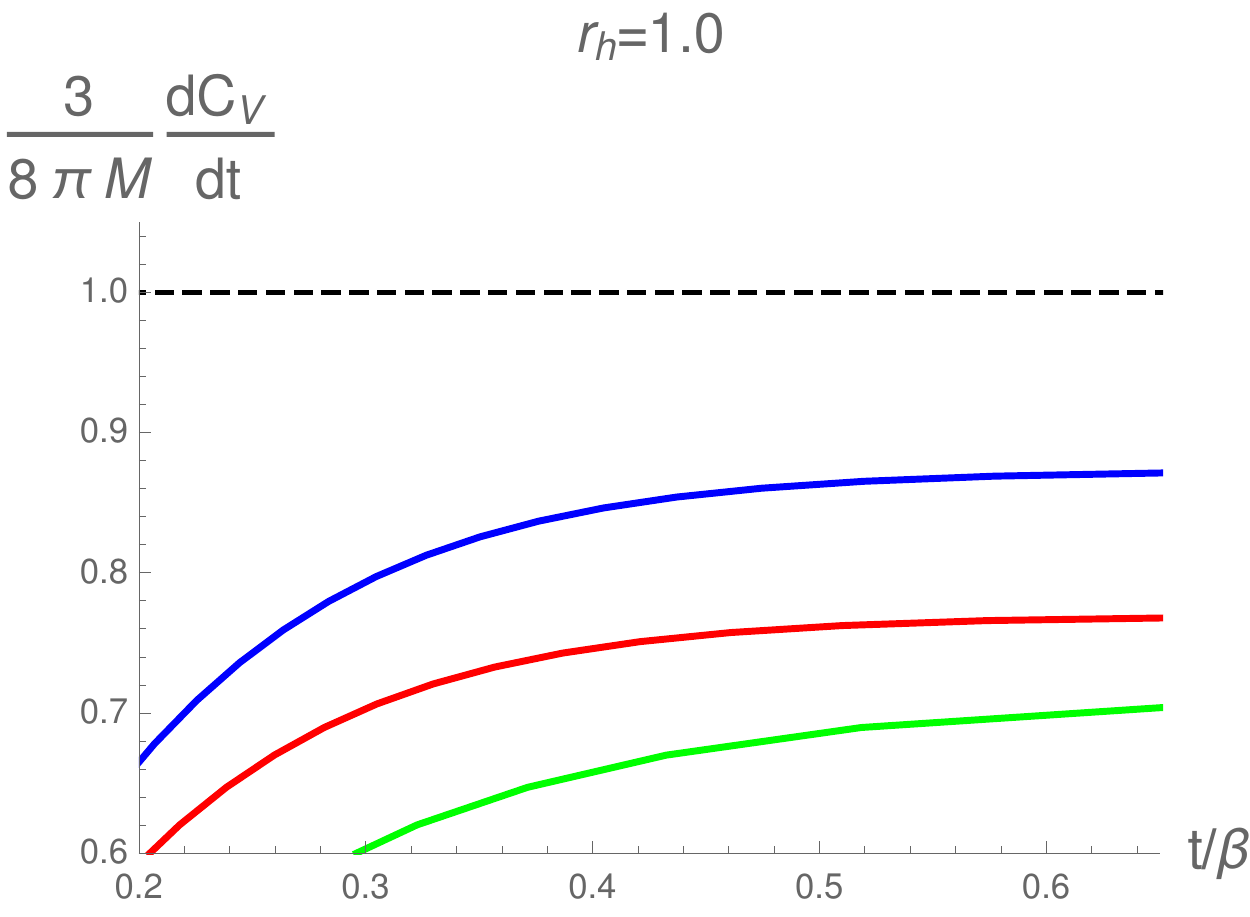}
	}
	\subfigure[]{
		\includegraphics[scale=0.5]{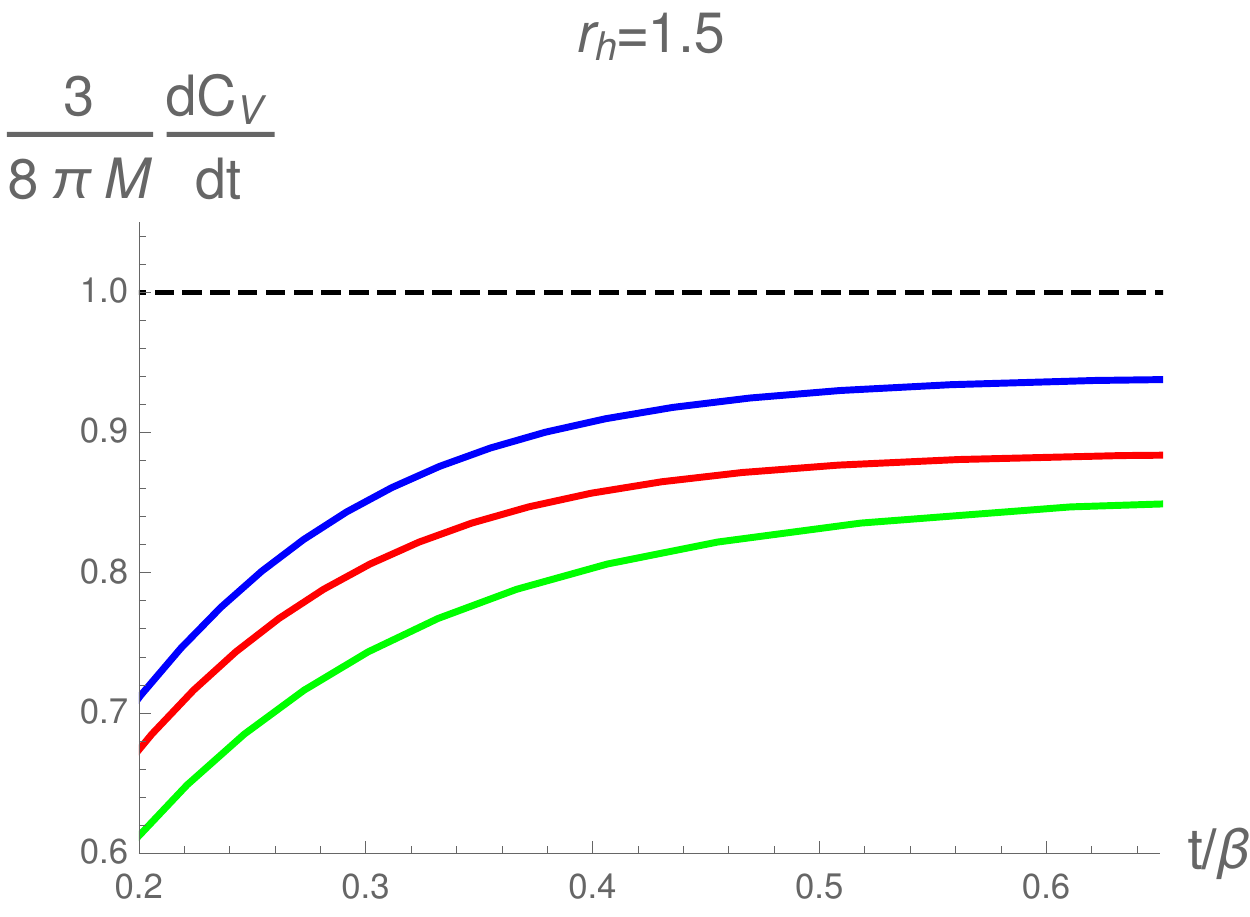}
	}
	\subfigure[]{
		\includegraphics[scale=0.5]{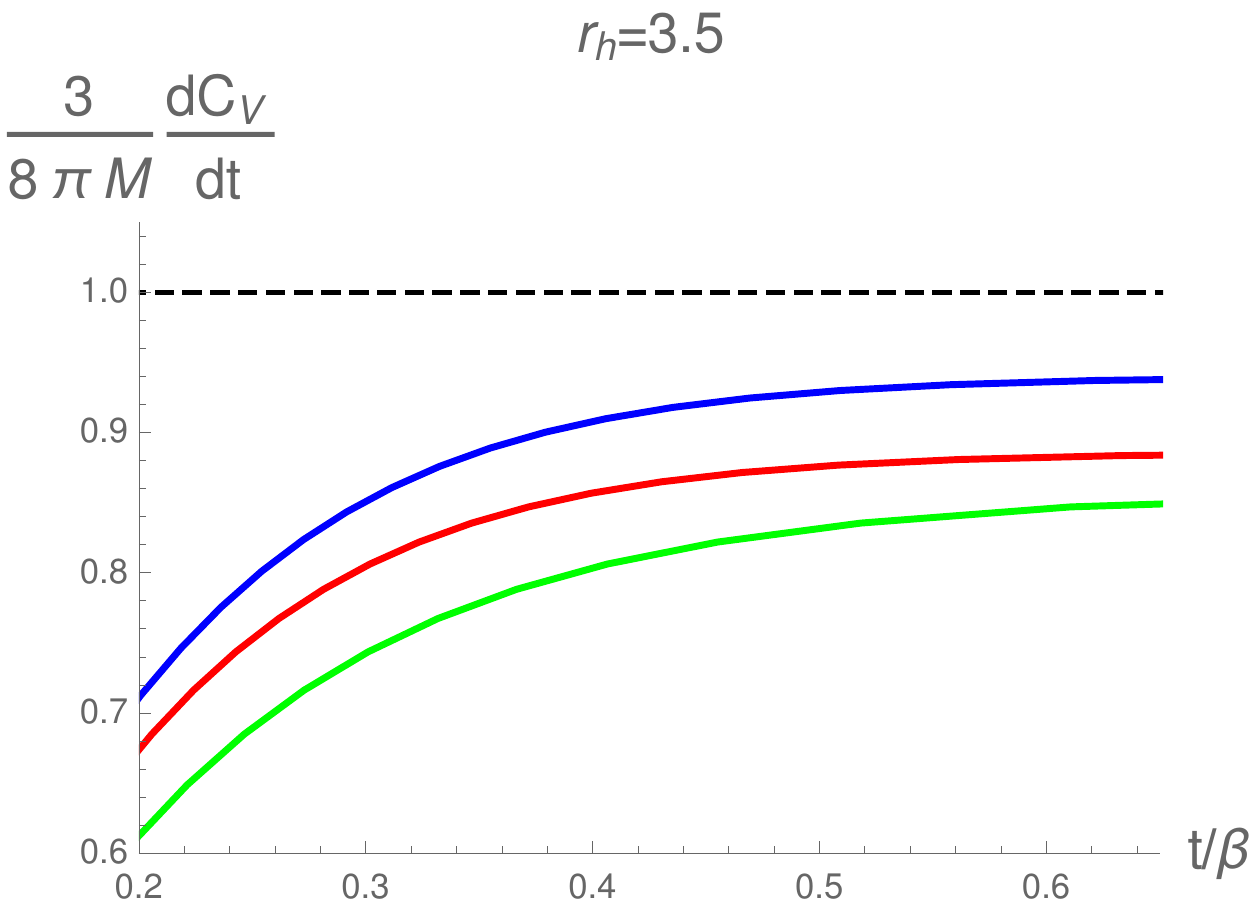}
	}
	\caption{$\frac{dC_V}{dt}$ as a function of $t/\beta$ for different values of the horizon radius $r_h$. Here we have used $d=4$ and $n=2$, and as before the blue curves indicate $a=0.05$, the red curves $a=0.1$ and the green curves $a=0.5$.}
	\label{fig:dCVdtd4n2}
\end{figure}

We may also plot full time dependence of the rate of change of complexity to investigate how it approach to its late-time value. With the expression for $t$ (eq. (\ref{boundarytime})) in hand, we now have all the ingredients necessary to plot the rate of change of complexity against the dimensionless quantity $t/\beta$ for different values of $r_h$ and $a$. The results are shown in Figs. \ref{fig:dCVdtd4n1} and \ref{fig:dCVdtd4n2}, and can be summarized as follows. Firstly, we find that for a fixed $t/\beta$ the magnitude of $\frac{dC_V}{dt}$ increases with the horizon radius. This result can be further mapped to our previous observation that $\frac{dC_V}{dt}$ approaches or saturates to its asymptotic value more early for larger size black holes. The comparison between blue, red and green curves in any of the figures further reveals that increase of $a$ leads to a decrease in the value of $\frac{dC_V}{dt}$. Interestingly, as opposite to the CA case, the change in the value of $n$ does not lead to any qualitative change in the behaviour of CV. Moreover, as we will see shortly, this behavior remains the same even when other values of spacetime dimensions are considered.

\subsection{Case 2: $d=3$, $n=1,2$}

In this subsection, we present the results of the rate of change of complexity for $d=3$, which should lend further credence to the conclusions drawn in the previous subsection. As in the case of $d=4$, we first discuss the late time behaviour of $\frac{dC_V}{dt}$ and then discuss its full time dependence.  Since most of the results are same as in the previous subsection we will therefore be very brief here.

\begin{figure}
	\begin{tabular}{cc}
		\includegraphics[scale=0.5]{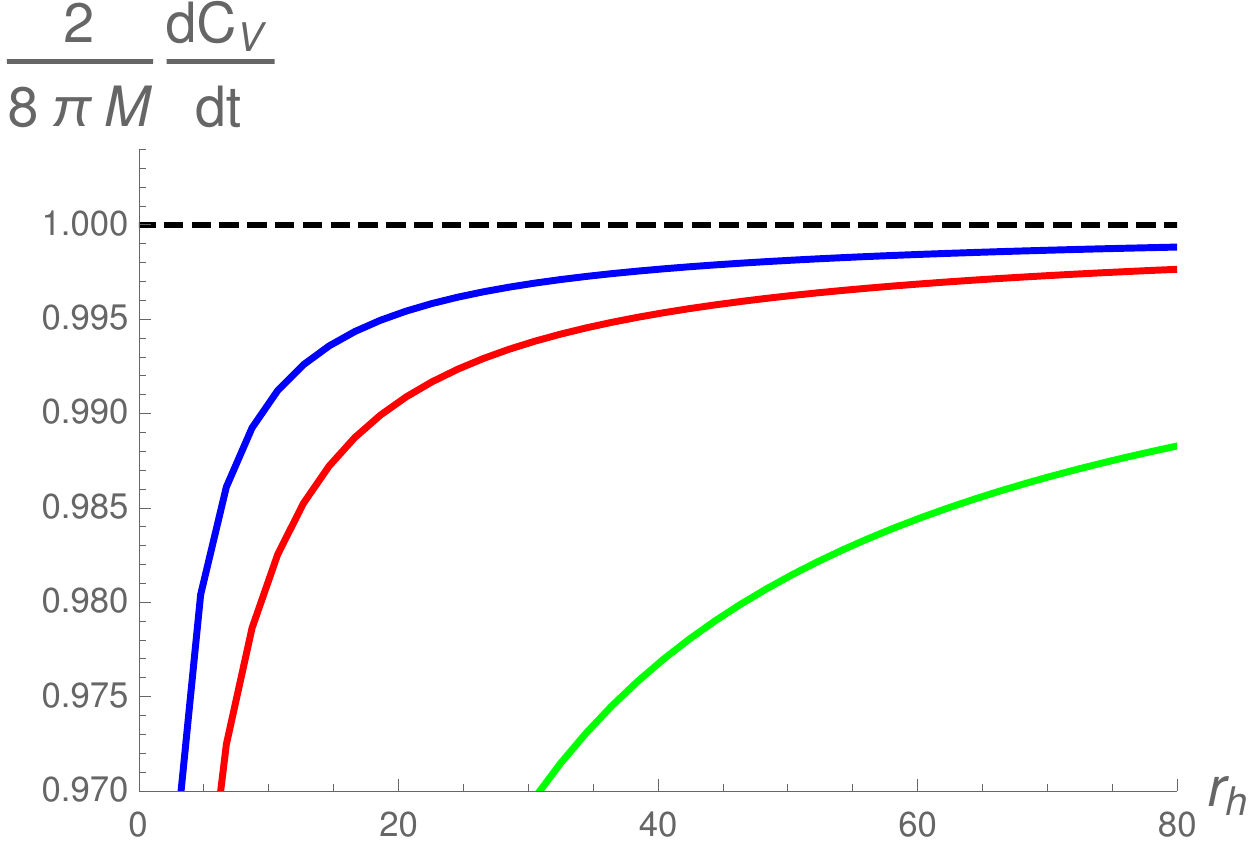}
		\includegraphics[scale=0.5]{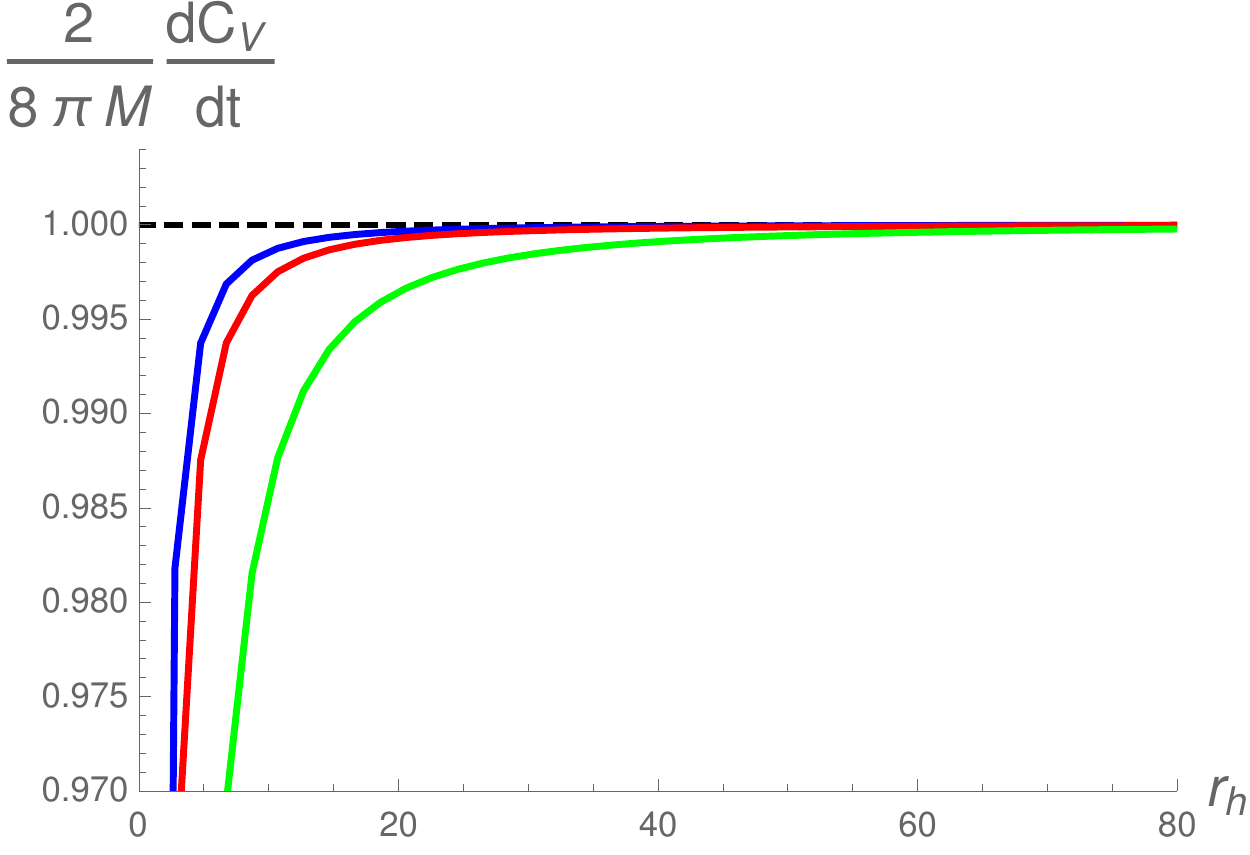}\\
		a & b
	\end{tabular}
	\caption{Panel (a) shows the variation of $\frac{dC_V}{dt}$ vs $r_h$ for $d=3$ for $n=1$ in the late time limit and (b) shows the same variation for $d=3$ and $n=2$. In both cases, the dotted line represents the ratio $\frac{2}{8\pi M}\frac{dC_V}{dt}$. The blue curves indicate $a=0.05$, the red curves $a=0.1$ and the green curves $a=0.5$.}
	\label{fig:dCVdtd3}
\end{figure}

\begin{figure}
	\subfigure[]{
		\includegraphics[scale=0.4]{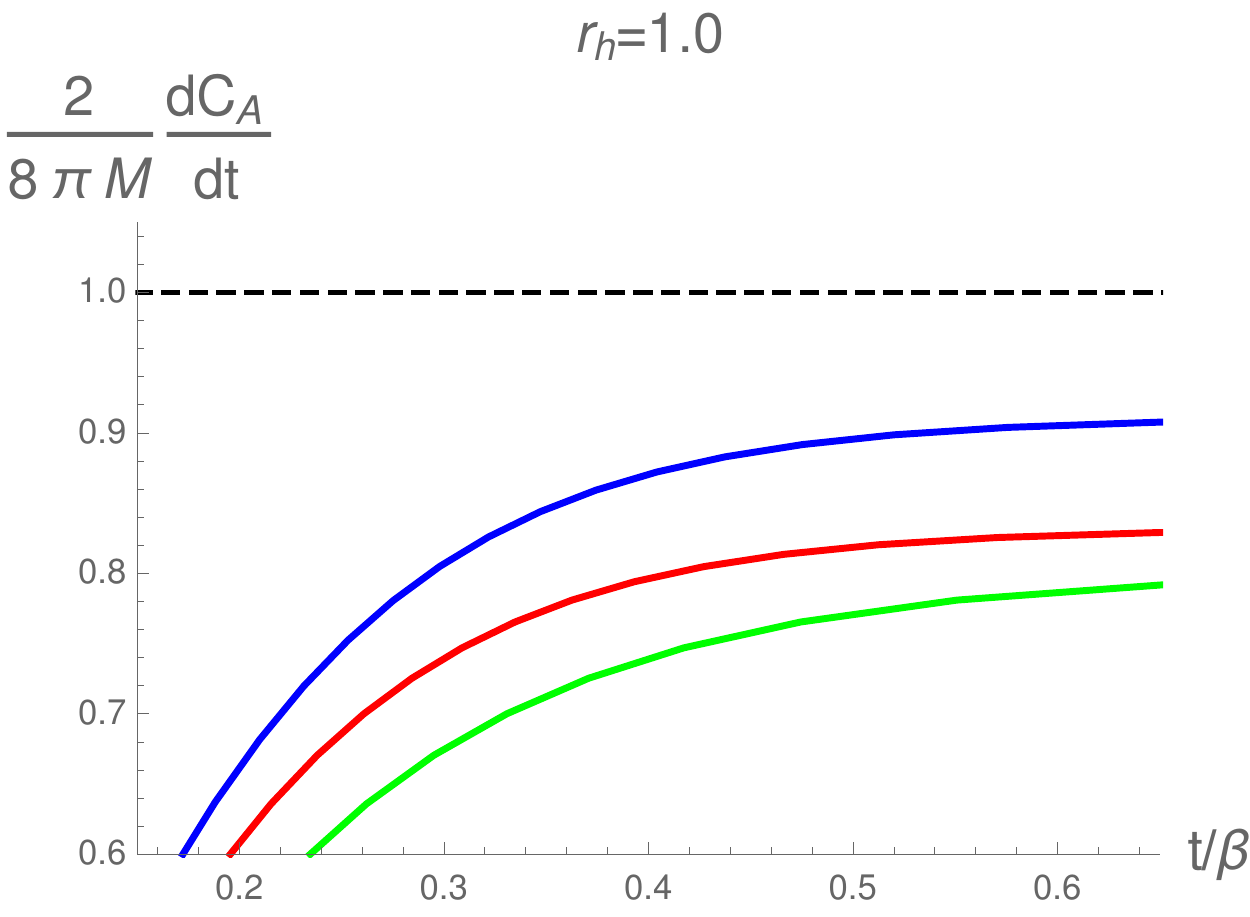}
	}
	\subfigure[]{
		\includegraphics[scale=0.4]{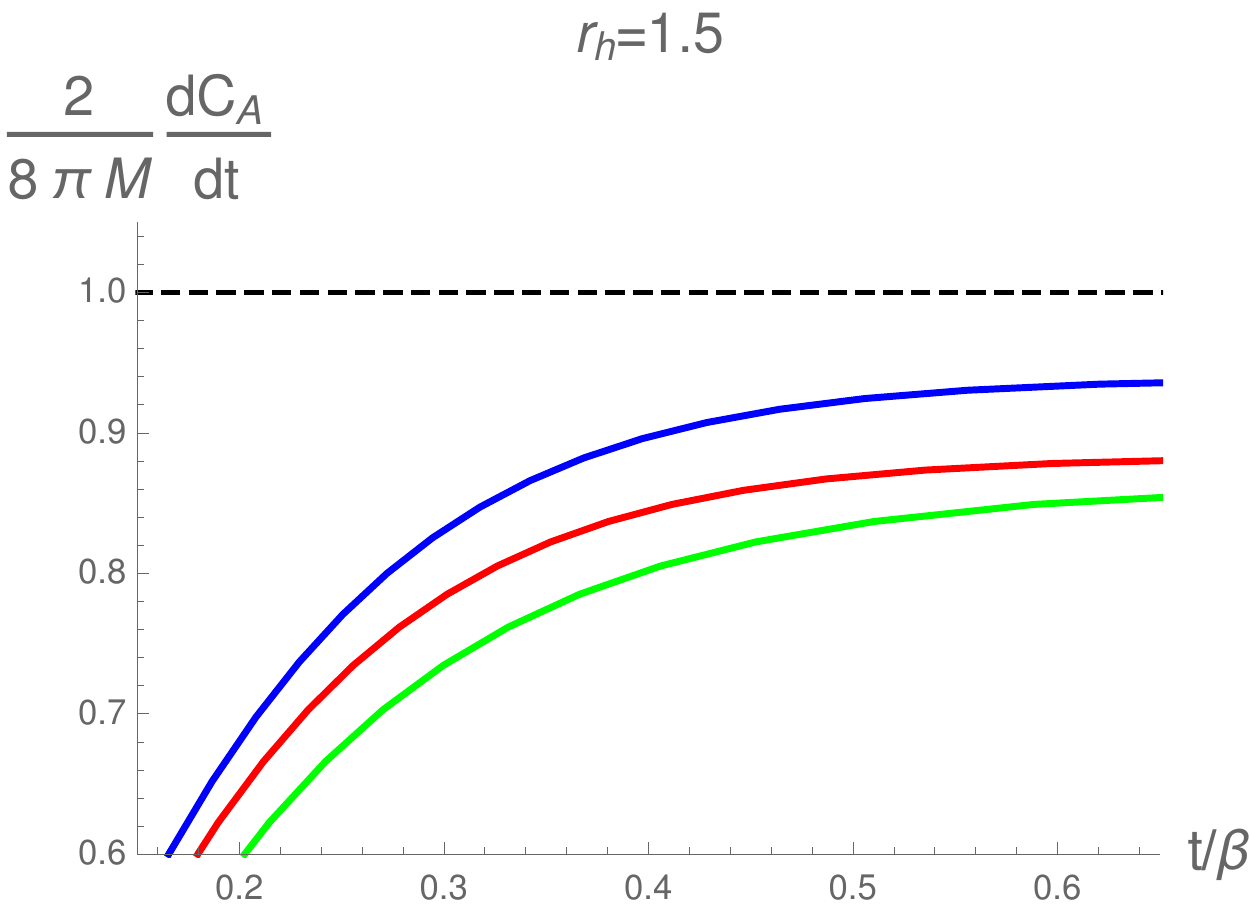}
	}
	\subfigure[]{
		\includegraphics[scale=0.4]{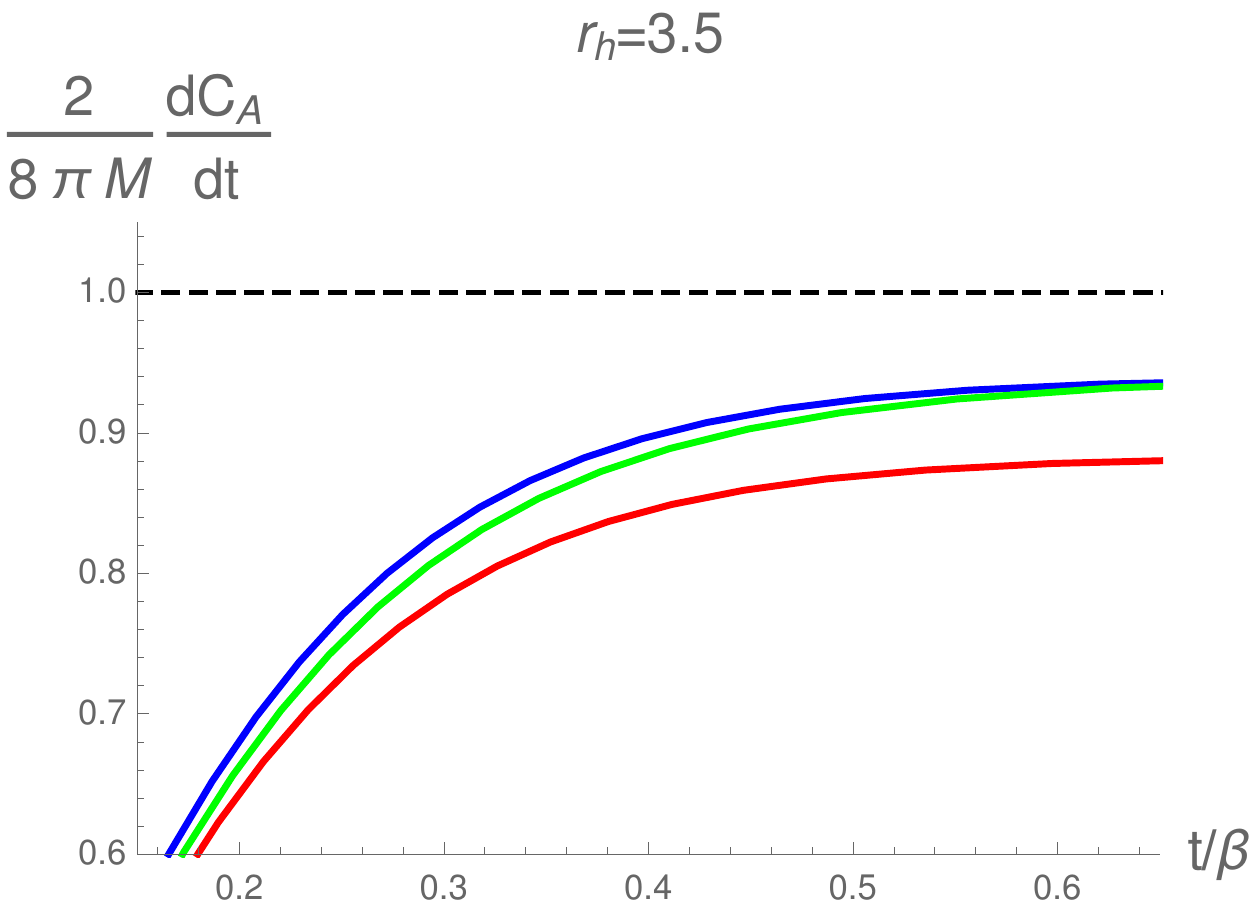}
	}
	\caption{$\frac{dC_V}{dt}$ is plotted against $t/\beta$ for different values of the horizon radius $r_h$. Here we have used $d=3$ and $n=1$, and as before, the blue curves indicate $a=0.05$, the red curves have $a=0.1$ and the green curves $a=0.5$}
	\label{fig:dCVdtd3n1}
\end{figure}

Our results for the late time behaviour of $\frac{dC_A}{dt}$ are shown in Fig. \ref{fig:dCVdtd3}.  As before, we see that the approach to the asymptotic value $\frac{8\pi M}{2}$ occurs at lower $r_h$ for increasing $n$. This is in line with the late time behaviour observed in $d=4$. We like to remind that for $n=2$ our model exhibits a Hawking-Page type phase transition, but similar to the example considered in \cite{Carmi:2017jqz}, the quantity $\frac{dC_V}{dt}$ is not sensitive to it. The numerical results for the general time dependence of $\frac{dC_V}{dt}$ are shown in Fig. \ref{fig:dCVdtd3n1}. For brevity, we present results only for $n=1$, as the results for $n=2$ are qualitatively similar. We again find the same qualitative features as in $d=4$. In particular, we again find that increase in the value of $a$ leads to an overall decrease in the value of $\frac{dC_V}{dt}$.

In summary to the results of this section, we may conclude that the CV proposal applied to this Einstein-dilaton model yields results that are consistent with those obtained previously in the literature. Although we do have two additional parameters in our model $(n,a)$, however their effects do not lead to a significant departure from the standard results of $\frac{dC_V}{dt}$.

\section{Summary}
\label{summary}

In this paper, we have studied the effects of non-trivial dilation on the holographic complexity using the CA and CV proposals by considering Einstein-dilaton action in the gravity side. We first obtained the gravity solution analytically in terms of an arbitrary scale function $A(r)$ in all spacetime dimension and then studied the time evolution of the complexity using CA and CV proposals. We found an explicit violation of the Lloyd bound using the CA proposal. In particular, we found that although the complexity is still proportional to the mass of the black hole however the proportionality factor is different in different spacetime dimensions, causing bound violation both in early as well as in late times. Moreover, we found that the deviation from the Lloyd bound is smaller for higher spacetime dimensions. Our work therefore provides another example in the growing list of works where the Lloyd bound in holographic theories can be explicitly violated. We moreover found that the additional parameters of our model, namely $n$ and $a$, can further modify the time evolution of $\frac{dC_A}{dt}$. In particular, $n$ and $a$ can change the early time behavior of $\frac{dC_A}{dt}$ without changing its late time structure. This is interesting because, as we mentioned in section 2, both $n$ and $a$ can drastically change the thermodynamics of the gravity system, especially the Hawking/Page type thermal $AdS$/black hole phase transition that appears even for the planar horizons in our model. It is interesting to note at this point that the holographic proposal for the entanglement entropy has been extensively used to probe the black hole phase transition and its thermodynamics \cite{Dudal:2017max,Johnson:2013dka,Dey:2015ytd}, and it might be interesting to investigate the same using holographic complexity proposals. Some work on this has already been initiated \cite{Zhang:2017nth,Zangeneh:2017tub,Zhang,Ghodrati:2018hss} \footnote{During preparation of this manuscript, we came to be aware of \cite{Zhang} which treats subregion complexity in the same model considered in this paper.}  and it would be fruitful to undertake the same investigations in a wider variety of models. The time evolution of $\frac{dC_V}{dt}$ in our model, as opposed to the CA proposal, however does not lead to any violation of the results found in \cite{Stanford:2014jda,Carmi:2017jqz}.

There are various directions to extend our work. In particular, we can include the chemical potential and background magnetic field and study their effects on the time evolution of the complexity. For this purpose we have to consider the Einstein-dilaton-Maxwell (EMD) action in the gravity side and solve a system of second order coupled differential equations, which may appear bit difficult to solve simultaneously. However, interestingly the potential reconstruction method can be used to solve EMD model as well \cite{Dudal:2017max,Dudal:2018ztm,Yang:2015aia}. Indeed, it has been observed that the chemical potential and dilaton field do leave non-trivial imprints on the structure of holographic complexity \cite{Carmi:2017jqz,Swingle:2017zcd}. Therefore it is interesting to study their effects in our model as well. Moreover, the effects of a background magnetic field on holographic complexity has not been studied yet. We hope to come back to these issues in near future.

\section*{Acknowledgments}
The work of S.~M.~is supported by the Department of Science
and Technology, Government of India under the Grant Agreement number IFA17-PH207 (INSPIRE Faculty Award).

\appendix
\renewcommand{\theequation}{\thesection.\arabic{equation}}
\addcontentsline{toc}{section}{Appendix}
\section{Appendix A: Black hole mass from Ashtekar-Magnon-Das(AMD) prescription}

 We refer the readers to \cite{Ashtekar:1999jx} for detailed discussion on Ashtekar-Magnon-Das (AMD) prescription and here we simply state the relevant equation needed for our analysis. In AMD prescription, the conserved quantity $C[K]$ associated with a Killing field
$K$ in an asymptotically AdS spacetime is given as,
\begin{equation}
\mathcal{C}[K]=\frac{1}{8 \pi (d-2)G_{d+1}} \oint \tilde{\epsilon}^{\mu}_{ \ \nu}  K^{\nu} d\tilde{\varSigma}_{\mu} \,.
\label{charge}
\end{equation}
where $\tilde{\epsilon}^{\mu}_{ \ \nu}=\Omega^{d-2} \tilde{n}^\rho \tilde{n}^\sigma \tilde{C}^{\mu}_{\ \ \rho \nu \sigma}$, $\Omega=1/r$ and $K^{\nu}$ is the conformal killing vector field.
$\tilde{C}^{\mu}_{\ \ \rho \nu \sigma}$ is the Weyl tensor constructed from $\tilde{ds^2}=\Omega^2 ds^2 $  and
$\tilde{n}^\rho$ is the unit normal vector to constant $\Omega$ surface.
$d\tilde{\varSigma}^{\mu}$ is the $d-1$ dimensional area element of the transverse section of the AdS  boundary. For a
timelike killing vector, we get the following expression for the conserved mass in our case
\begin{equation}
\mathcal{C}[K]=M=\frac{V_{d-1}}{8 \pi (d-2) G_{d+1}} \Omega^{2-d} (\tilde{n}^{\Omega})^{2}\tilde{C}^{t}_{\ \ \Omega t \Omega} \,.
\end{equation}
substituting the expression of $\tilde{C}^{t}_{\ \ \Omega t \Omega}$ and switching back to $r=1/\Omega$ coordinate, we get the following expression for the black hole mass $M$,
\begin{eqnarray}
M &=& -\frac{V_{d-1}}{8 \pi (d-2) G_{d+1}} \left[\frac{d-2}{d}r^{d+1}g'(r)+\frac{d-2}{2d} r^{d+2}g''(r) \right] \ , \nonumber \\
  &=& -\frac{V_{d-1}}{8 \pi G_{d+1}} \frac{r^{d+1}}{d} \left[ g'(r)+\frac{1}{2} r g''(r)   \right]
  \label{mass}
\end{eqnarray}
now substituting the expressions of $A(r)$ and $g(r)$ in eq. (\ref{mass}), we can calculate the black hole mass in Einstein-dilaton gravity in any dimension.

Generally due to the presence of matter fields, the behavior of metric at the asymptotic boundary can be different from that arising from pure
gravity. In particular, if the matter fields do not fall off sufficiently fast at the asymptotic boundary then it can lead to a different asymptotic behavior of the metric, which further can lead to a different expression of the conserved charges, for example see \cite{Henneaux:2004zi}. Although, in our gravity model, we too have a non-trivial profile of the scalar field which backreacts to the spacetime geometry, however importantly it does not modifies the asymptotic structure of the metric. In particular, the metric coefficients have the same order of falloffs at the boundary even with the scalar field. Therefore, the usual expressions of the conserved charges in asymptotic AdS spaces, like the one in eq. (\ref{charge}), remains the same.  For detailed discussion on different methods to calculate conserved charges in asymptotic AdS spaces and relation between them, see \cite{Hollands:2005wt}. In any case we also have checked that the holographic renormalisation procedure gives the same expression for the black hole mass, albeit with a constant offset, as considered at various places in this paper.

\section{Appendix B: Derivation of the counterterm}

In this appendix, we derive the form of the counterterm that removes the ambiguity related to the parametrization of the null normals. Briefly, this term is intended to remove the $c, \bar{c}$-dependence in the joint term in the CA calculation of complexity (see eq. \ref{joint}). According to \cite{Lehner:2016vdi}, the term that needs to be added to the action our results to be independent of the choice of normalization of the null normals is,
\begin{equation}
S_{ct} = \frac{V_{d-1}}{8\pi G_{d+1}}\int_{\Sigma} d\lambda d^{d-1} y \sqrt{\gamma} \Theta \log(\ell_{ct}|\Theta|)
\label{counterterm}
\end{equation}
where $\gamma$ refers to the metric on the null generators and $\lambda$ is a parameter for the generators of the null boundary. The quantity $\ell_{ct}$ which appears in the prescription for the counterterm is an arbitrary length scale which is related to the ambiguity of normalization of the null normals. $\Theta=\partial_{\lambda}\log{\sqrt{\gamma}}$ is the expansion scalar of the null boundary generators. Recalling that, in the main body of the paper we have chosen to parametrize the generators of the null hypersurfaces to be affinely parametrized, we set $\lambda = r/\alpha$ for which it may be checked,
\begin{equation}
k^{\mu} \nabla_{\mu} k_{\nu} = 0
\end{equation}
so that $\lambda$ is an affine parameter. With this parametrization, the total counterterm contribution takes the form,
\begin{eqnarray}
S_{ct} = \frac{3 V_3}{4\pi G_5}\int_{0}^{r_{max}} dr r^2 e^{-\frac{3 a}{r}} \left(\frac{a}{r}+1\right) \log \left(\frac{3 \alpha  \left(\frac{a}{r}+1\right) \ell _{ct}}{r}\right) + \nonumber \\ \frac{3 V_3}{4\pi G_5} \int_{r_m}^{r_{max}} dr r^2 e^{-\frac{3 a}{r}} \left(\frac{a}{r}+1\right) \log \left(\frac{3 \alpha  \left(\frac{a}{r}+1\right) \ell _{ct}}{r}\right)
\end{eqnarray}
In the above expression for the counterterm, we have set $d=4, n=1$ for which the integration can be analytically performed. Also, we have set $c, \bar{c}=\alpha$ in the expressions for simplicity. The limits of the above integrals is explained by noting that we have to add counterterms for each of the future and past null boundaries (see Fig.~\ref{fig:wdw}). Performing the integral we have the following expression,
\begin{eqnarray}
 S_{ct} = (\#) \rvert_{r=r_{max}} - (\#) \rvert_{r=0} + (\#) \rvert_{r=r_{max}} - (\#) \rvert_{r=r_m} \nonumber \\ \text{where,} \ \ \   \# = \frac{1}{9} r e^{-\frac{3 a}{r}} \left(-3 a^2+3 r^2 \log
\left(\frac{3 \alpha  (a+r) \ell _{ct}}{r^2}\right)+r^2\right) \nonumber \\
+ \frac{1}{3} a^3 \left( Ei\left(-\frac{3 (a+r)}{r}\right)-4 Ei\left(-\frac{3 a}{r}\right)\right)
\end{eqnarray}
Since we are interested in the time dependence of the counterterm, we need only consider the term involving $r_m(t)$. Accordingly, the total time derivative of the joint term (including the counterterm) is,
\begin{eqnarray}
 \frac{dS_{jnt}}{dt} &=& \frac{V_3}{16 \pi G_5} \left( r_m^3 e^{-\frac{3 a}{r_m}} g(r_m) \left(3 (a+r_m) \left(4 \log \left(\frac{3 (a+r_m) \ell _{ct}}{r_m^2}\right)+  \log
(G(r_m))\right)\right) \right) \nonumber \\ &&  + \frac{V_3}{16 \pi G_5} \frac{r_m^5 e^{-\frac{3 a}{r_m}}  g(r_m)  G'(r_m)}{G(r_m)}
\end{eqnarray}
As can be observed from the above expression, the $\alpha$-dependence is removed from the joint term. Also, since $g(r_h)=0$ and $r_m \rightarrow r_h$ at late times, it is clear that the counterterm does not contribute at late times and hence would not alter the late time value of the rate of change of complexity. Although the counterterm in eq. (\ref{counterterm}) does not change our results for $\frac{dC_A}{dt}$, it may effect $\frac{dC_A}{dt}$ behaviour in the case of shock wave geometries \cite{Chapman:2018dem,Chapman:2018lsv}.

\end{document}